\documentclass[fleqn,usenatbib]{mnras}

\usepackage{newtxtext,newtxmath}
\usepackage{mathtools}
\usepackage{xspace}

\usepackage[T1]{fontenc}

\DeclareRobustCommand{\VAN}[3]{#2}
\let\VANthebibliography\thebibliography
\def\thebibliography{\DeclareRobustCommand{\VAN}[3]{##3}\VANthebibliography}

\usepackage{graphicx}	
\usepackage{amsmath}	
\usepackage{threeparttable} 

\begingroup
\catcode`\_=\active
\gdef_#1{\ensuremath{\sb{\mathrm{#1}}}}
\endgroup
\mathcode`\_=\string"8000
\catcode`\_=12

\newcommand{\cc}{cm$^{-3}$}                          

\newcommand{\gadget}{{\sc gadget}-2\xspace}                
\newcommand{\hcc}{H~cm$^{-3}$}                       

\newcommand{\hm}{H$_2$}                              
\newcommand{\jwst}{\textit{JWST}\xspace}
\newcommand{\msun}{{\rm M}_{\odot}}                  
\newcommand{\nh}{n_{H}}                          
\newcommand{\ramses}{{\sc ramses}\xspace}                
\newcommand{\ramsesrt}{{\sc ramses-rt}\xspace}           
\newcommand{\subi}{\textit{i}}
\newcommand{\subj}{\textit{j}}

\newcommand{\outward}{Fiducial-Outward\xspace}      
\newcommand{\inward}{Fiducial-Inward\xspace}      
\newcommand{\non}{Fiducial-Neutral\xspace}      

\title[Outward migration of Population~III stars]{On the origin of outward migration of Population~III stars}

\author[J. Park, M. Ricotti and K. Sugimura]{
Jongwon Park,$^{1}$\thanks{E-mail: jwpark@umd.edu} 
Massimo Ricotti,$^{1}$\thanks{E-mail: ricotti@umd.edu}
and Kazuyuki Sugimura$^{2}$
\\
$^{1}$Department of Astronomy, University of Maryland, College Park, MD 20742, USA\\
$^{2}$Faculty of Science, Hokkaido University,
Sapporo, Hokkaido 060-0810, Japan\\
}

\date{Accepted XXX. Received YYY; in original form ZZZ}

\pubyear{2023}

\begin{document}
\label{firstpage}
\pagerange{\pageref{firstpage}--\pageref{lastpage}}
\maketitle

\begin{abstract}
Outward migration of massive binary stars or black holes in their circumbinary disc is often observed in simulations and it is key to the formation of wide black hole binaries. Using numerical simulations of Population~III (Pop~III) star formation, we study the angular momentum of Pop~III binaries and the torques between stars and gas discs to understand the origin of outward migration and high ellipticity. The outward migration of protostars is produced by gravitational torques exerted on them by their circumstellar minidiscs. The minidiscs, on the other hand, migrate outward mainly by gaining angular momentum by accreting gas from the circumbinary disc. The angular momentum transfer is most efficient for rapidly accreting equal-mass binaries, and weaker when the secondary mass is small or the massive companion evaporates the gas disc via radiative feedback. We conclude that outward migration and the formation of wide equal-mass massive binaries is common in metal-free/metal-poor star formation, mainly driven by their large accretion rates. We expect that the lower gas temperature and accretion rates in metal-enriched circumstellar discs would lead more often to inward migration and closer binary separations. We also observe inward migration for smaller mass Pop~III protostars/fragments, leading to the rapid merging of sink particles and likely the formation of close binary black holes that, however, reach separations below the resolution of our simulations. We discuss the implications that Pop~III separations and ellipticity may have on the interpretation that gravitational wave signals from merging intermediate-mass black holes come from Pop~III remnants.
\end{abstract}




\section{Introduction}

In recent years understanding the formation and properties of metal-free first stars (Population~III or Pop~III stars) has gained renewed impetus thanks to the discovery of gravitational wave (GW) emission from merging intermediate-mass black holes (IMBHs) and the successful launch of the \textit{James Webb Space Telescope} (\jwst). The detection by VIRGO and LIGO of signals from merging black hole binaries (BHBs) with individual masses $\sim 30~\msun$ \citep{abbott2016,abbott2017} indicate that they might be the remnants of metal-poor or metal-free star formation \citep{hartwig2016,liu2020}. The \textit{Hubble Space Telescope} and the \jwst have detected a few high-$z$ lensed stellar objects at $z\sim 5-6$ \citep[e.g.,][]{welch_highly_2022, welch_jwst_2022}, thereby suggesting detection of Pop~III stars is possible at high magnification. The search of Pop~III stars with \jwst is actively ongoing and strongly-lensed extremely metal-poor small mass star clusters ($10^4$~M$_\odot$ with $Z \sim 10^{-3}$~Z$_\odot$, possibly with a top-heavy IMF) at $z\sim 6$ have already been observed \citep{vanzella2023}. Remarkably, compact objects with similar properties have been predicted to exist in simulations of the first galaxies \citep{garcia2023}. Finally, massive Pop~III stars may explode as pair-instability supernovae \citep{heger2002}, and these extreme events may be detected by the \textit{JWST} and in the near future by the \textit{Nancy Grace Roman Space Telescope} \citep{whalen2014}. Given this context, understanding the formation and properties of Pop~III stars and more generally the fragmentation and evolution of metal-poor gas clouds and circumstellar discs is especially timely and well-motivated.

Among the many properties of Pop~III stars, their multiplicity is still poorly understood, despite its importance. There is a general consensus that the fragmentation of a metal-free gas disc and the subsequent formation of Pop~III binaries is a common occurrence (\citealp{machida2008}; \citealp*{stacy2010}; \citealp{clark2011,susa2013}; \citealp*{susa2014}; \citealp{hosokawa2016,sugimura2020}). Once they form, these binaries can migrate either inward or outward. Inward migration may lead to the formation of close BHBs or to mergers between Pop~III protostars. In the latter case, this enhances the growth of the primary star \citep{greif2012,hosokawa2016} and reduces the multiplicity of the Pop~III star system, thereby reshaping the Pop~III initial mass function (IMF). Inward migrations appear commonly in several numerical simulations \citep{greif2012,stacy2013,hirano2017,chon2019} but outward migrations have also been observed \citep{greif2011,greif2012,stacy2013,chon2019,sugimura2020}. 

A series of papers \citep[][hereafter, Papar~I, II, and III, respectively]{park2021a,park2021b,park2023} found that the formation of wide hierarchical binaries via outward migration, often with relatively large eccentricities, is the most typical outcome for massive Pop~III stars (semimajor axis, $a_{b} \sim$ a few 1,000~AU to 10,000~AU). This is the fourth paper in the series focusing on understanding the physical mechanisms driving outward migration and exploring the origin of high eccentricities.
In Paper~III, we underscored their importance in forming wide Pop~III BHBs because even these may lead to BH mergers and GW emission through two different dynamical channels: dynamical hardening \citep{liu2020} or orbital excitation \citep{michaely2019,michaely2020}. 

Due to its ubiquity and importance, there have been several previous studies aimed at understanding the origin of outward migration in circumbinary discs. Several authors studied ejections via N-body processes \citep{greif2011,greif2012,stacy2013} or gas accretion \citep[][CH19 hereafter]{chon2019}. This latter study focused on the evolution at small scales (tens of AUs) while Pop~III binaries typically reach separations up to 10,000~AU over a timescale of 50-100~kyr (\citealp{sugimura2020}; Paper~II; Paper~III). These previous works suggested the accretion of high angular momentum gas as the dominant mechanism for outward migration. In Paper~II, although we lacked a quantitative analysis, we discussed the possible connection between outward migration and gas accretion. The migration rate and separation of wide binaries were found to depend on the intensity of an external X-ray background in the following way. An X-ray background enhances the \hm\ cooling and thus lowers the gas temperature. The reduced sound speed $c_{s}$ leads to a lower accretion rate due to its $c_{s}^3$ dependence. In an X-ray, therefore, protostars grow more slowly and accrete less high-angular momentum gas, and thus stars have lower masses and the extent of their outward migration is reduced. For this reason, the maximum separation of wide binaries tends to be smaller in an intense X-ray radiation background (see Fig.~8 of Paper~II). This correlation, however, was less evident in Paper~III which differs from our previous work in that it included a full treatment of radiative feedback from accreting protostars. 

In this work, we analyse the simulations presented in Paper~III to study in detail the physical mechanism that induces outward migrations and ultimately produces the large separations of hierarchical Pop~III binaries. As discussed above, previous works on stars and black hole binaries had some disagreement on whether the main mechanism for migration is gravitational torques or the accretion of high angular momentum gas. We aim at clarifying this issue and analyse the role of radiative feedback that is likely responsible for the somewhat different results between Paper~II and Paper~III regarding migration. The ultimate goal is to determine what are the unique conditions in Pop~III protostellar discs that lead to migration and eccentric orbits, and why this behaviour is not observed in simulations of binaries formed in protostellar discs with solar metallicity \citep{he2023}.

Finally, a methodology note: There is a large body of work focusing on understanding the physics of star and black hole binaries migration in gas discs (e.g., \citealp*{tang2017}; \citealp*{munoz2019}; \citealp{munoz2020}; \citealp*{dempsey2020}; \citealp*{dempsey2021}; \citealp{dittmann2021,dittmann2022}). This field of research typically adopts idealised initial and boundary conditions and neglects radiative feedback effects. For these reasons, these works can typically achieve higher resolution, evolve the systems for hundreds of orbits, and perform a systematic exploration of disc parameters and numerical experiments. This paper is instead a follow-up in a series of papers on Pop~III stars formation from cosmological initial and boundary conditions (see Paper~I, II, and III). We present one of the most detailed and in-depth analysis of the physics of Pop~III binary migration when compared to previous Pop~III literature but the present paper has a different focus and it is hard to compare to the results from idealised discs simulations, even though it has a strong cross-section with these papers. Nevertheless, given the caveat mentioned above, when possible we will highlight and discuss the connections between our results on Pop~III stars and the rich literature on binary migration in idealised gas discs.

This paper is organised as follows. In Section~\ref{sec:method} we summarise the method and simulation focusing on the features relevant to this work. In Section~\ref{sec:result1}, we explore the origin of outward migration. In Section~\ref{sec:result2}, we discuss binaries without outward migration to better understand the outward cases. In Section~\ref{sec:discussion} and \ref{sec:summary}, we discuss the implication of our findings and summarise the main conclusions.


\section{Method}
\label{sec:method}

\subsection{Simulations}
\label{sec:sim}
\subsubsection{Overview}
We make use of Adaptive Mesh Refinement (AMR) code \ramsesrt\ \citep{teyssier2002,rosdahl2013} to simulate the formation of Pop~III stars in metal-free gas discs extracted from minihaloes in cosmological simulations. The non-viscous gas motion is described by solving the Euler equations using a second-order Godunov method and an approximate solution of the radiative transfer equations for UV radiation emitted from massive protostars is included and coupled to primordial chemistry (\citealp{rosdahl2013}; Paper~I; Paper~III). Motivated by early analytic results presented in \cite{Ricotti2016}, in addition to radiative feedback from protostars, we include external radiation backgrounds in the Lyman-Werner (LW) bands and X-rays. Time-dependent primordial gas chemistry and relevant cooling/heating processes for gas densities up to $10^{14}$~cm$^{-3}$ are included. In this paper we analysed the same simulations run in Paper~III, hence we refer to that paper for details on the simulation methods and convergence tests (see Appendix in Paper~III). Here, for completeness, we only summarise some basic information on the simulations and the parameters of different runs.

We extract the central regions ($2$~pc) at the centre of two minihaloes in cosmological zoom-in simulations (see Fig.~2 of Paper~III) to create the initial conditions. These initial conditions include a metal-free gas cloud that has been irradiated by various X-ray and LW backgrounds. We performed 8 simulations and identified 14 binaries in them. Table~\ref{tab:sim} summarises the set of simulations and some properties of the binaries. Once a simulation begins, the gas cloud with peak hydrogen number density $\nh \sim 10^7$~\hcc\ contracts and flattens out to form a disc. As the cloud contracts and the central density increases, cells are refined from AMR level 7 to 15 with the Jeans refinement criterion $\lambda_{\subj} < N_{\subj} \Delta x$. Here, $\lambda_{\subj}$ is the local Jeans length in each gas cell, $\Delta x$ is the cell size, and $N_{\subj} = 16$ is the number of cells into which we resolve the Jeans length. The smallest cell size or the maximum spatial resolution is $\Delta x_{min} = 2~{\rm pc}/2^{15} = 12.6$~AU.
\begin{table*}
    \caption{Summary of simulations. From left to light, we show 1) binary name, 2) formation time, 3) final time, 4) separation at formation, 5) final separation, 6) eccentricity, 7) mass ratio, 8-9) mass of individual stars, and 10) type of migration.}
    \footnotesize
    \begin{threeparttable}
    	\centering
    	\label{tab:sim}
    	\begin{tabular}{ | l | c | c | c | c | c | c | c | c | c | }
		    \hline
            Label\tnote{a,b} & $t_{form}$ [kyr] & $t_{final}$ [kyr] \tnote{c} & $d_{form}$ [AU] & $d_{final}$ [AU] & $e$\tnote{d} & $q$\tnote{e} & $M_1 (\msun)$\tnote{f} & $M_2 (\msun)\tnote{f}$ & Migration\tnote{g} \\
		    \hline

            Run~A & & & & & & & \\
            S01-S02\tnote{$\star$} & $0.38$ & $68$ & $727$ & $7637$ & $0.79$ & $0.847$ & $183$ & $215$ & neutral \\
            \hline

            Run~B & & & & & & & \\
            S01-S04 & $0.30$ & $42$ & $1528$ & $1594$ & $0.46$ & $0.971$ & $88$ & $85$ & neutral \\
            \hline

            Run~C & & & & & & &\\
            S01-S02 & $0.22$ & $92$ & $584$ & $4927$ & $0.22$ & $0.943$ & $247$ & $249$ & outward \\
            - S01-S06 & $29$ & $92$ & $1544$ & $705$ & $0.63$ & $0.205$ & $247$ & $51$ & inward \\
            - S02-S12\tnote{\dag} & $51$ & $92$ & $1149$ & $484$ & $0.33$ & $0.132$ & $249$ & $33$ & inward \\
            \hline

            Run~D & & & & & & &\\
            S01-S02 & $0.31$ & $67$ & $492$ & $2910$ & $0.19$ & $0.901$ & $74$ & $82$ & outward \\
            \hline

            Run~E & & & & & & &\\
            S01-S02 & $0.30$ & $103$ & $498$ & $7623$ & $0.28$ & $0.971$ & $49$ & $51$ & outward \\
            \hline

            Run~F & & & & & & &\\
            S01-S03 & $0.83$ & $108$ & $664$ & $40982$ & $0.16$ & $0.800$ & $83$ & $67$ & outward \\
            - S01-S10\tnote{$\ast$} & $25$ & $108$ & $519$ & $5244$ & $0.34$ & $0.870$ & $83$ & $71$ & outward \\
            - S03-S05 & $15$ & $108$ & $426$ & $4240$ & $0.32$ & $0.840$ & $67$ & $56$ & outward \\
            \hline

            Run~G & & & & & & &\\
            S01-S03 & $1.6$ & $95$ & $1537$ & $43310$ & $0.27$ & $0.505$ & $73$ & $41$ & outward \\
            - S01-S02 & $0.31$ & $95$ & $558$ & $4846$ & $0.40$ & $0.971$ & $73$ & $75$ & outward \\
            - S03-S04 & $21$ & $95$ & $520$ & $6711$ & $0.61$ & $0.833$ & $41$ & $34$ & outward \\
            \hline

            Run~H & & & & & & &\\
            S01-S03 & $0.33$ & $33$ & $1799$ & $4405$ & $0.53$ & $0.775$ & $89$ & $114$ & outward \\
            \hline
	    \end{tabular}
	    \begin{tablenotes}
	        \item[a] Following the labels in Fig.~1 in Paper~III.
            \item[b] A hyphen before the name indicates the binary belongs to a hierarchical binary.
            \item[c] Final time of the simulation.
            \item[d] Maximum eccentricity.
            \item[e] $q=M_2/M_1$ if $M_1 \geq M_2$; $q=M_1/M_2$ if $M_1 < M_2$.
            \item[f] At $t_{final}$.
            \item[g] Outward if $d_{form} < d_{final}$; inward if $d_{form} > d_{final}$; neutral if $d_{form} \sim d_{final}$. An exception is Run~A where the orbit is kept highly eccentric.
            \item[$\star$] Representative case of non-migration binaries. Throughout the paper, it is referred to as ``\non''.
            \item[\dag] Representative case of inward migration binaries. Throughout the paper, it is referred to as ``\inward''.
            \item[$\ast$] Representative case of outward migration binaries. Throughout the paper, it is referred to as ``\outward''.
        \end{tablenotes}
    \end{threeparttable}
\end{table*}

\subsubsection{Sink prescription}
The flattened disc fragments and multiple blobs form in the disc or spiral structures. Using the built-in halo finder \citep{bleuler2015} we identify these dense blobs at every coarse timestep ($\sim 13$~yr) and assign a sink particle at the density peak of each blob if,
\begin{enumerate}
    \item $\nh$ at the peak exceeds $n_{sink} = 10^{12}$~\hcc,

    \item there is no other sink particle within $2r_{sink}$,

    \item the velocity field in the blob has a net negative divergence.
\end{enumerate}
A sink particle has the radius of $r_{sink} = 8\Delta x_{min} = 101$~AU. This resolution of sink is insufficient to resolve a close binary and therefore a single sink particle may represent multiple Pop~III stars. In this work, however, we assume that each sink particle describes a Pop~III star and use the terms `sink particle' and `star' interchangeably. If a sink particle mass exceeds $1~\msun$, it emits UV radiation. The luminosity is determined by the sink mass and accretion rate and is computed by interpolating the tabulated model of massive protostar by \citet{hosokawa2009} and \citet*{hosokawa2010}. We record the masses, positions, and velocities of the existing sink particles at every coarse timestep ($\sim 13$~yr).

If two sink particles are closer than $2r_{sink} = 202$~AU, they merge into one. The new sink particle is positioned at the centre-of-mass (CoM) of the two and momentum is conserved. We explored the impact of varying the sink radius in Appendix~\ref{app:rsink}. The acceleration of a sink particle by the gas or other sinks is softened assuming a smoothing-length $\varepsilon = 4\Delta x_{min} = 50.4$~AU. A difference of this study with respect to the more common approach of running idealised circumbinary disk (CBD) simulations is the adoption of softening the gravitational force between sink particles. Although typical CBD simulations ignore this modification, we keep this term in the force calculation for consistency with our previous cosmological simulations (Paper~I, II, and III). In Appendix~\ref{app:soft}, we show that the impact of force softening on our conclusions is negligible. The accretion of gas onto the sinks is performed by checking the gas density of each cell within $r_{sink}$. If a hydrogen number density $n_{H}$ exceeds the threshold value, the excess mass $(n-n_{sink})\Delta x^3$ is added to the sink and subtracted from the cell. The position and velocity of the centre-of-mass are conserved before and after the accretion. Recent numerical studies \citep{dempsey2020,dittmann2021} suggested that the sink particle methods of removing mass within a boundary may change the density profile unphysically and introduce an artificial (numerical) torque. We argue that this effect is negligible in our simulations because the time scales of our simulation differ from those of the above papers and strong radiative feedback by Pop~III stars is the dominant physical process shaping the gas density profile rather than gas removal within the sink. We discuss this topic in-depth in Appendix~\ref{app:accretion}. Sink particle motions are calculated using a second-order mid-point scheme which reduces to the leapfrog method with constant time step \citep{bleuler2014}. The tests performed by \citet{bleuler2014} showed that the energy and angular momentum are conserved in a Keplerian orbit. Note that, however, this test was performed to explore the impact of different force calculation methods, not N-body integrators. It is well known that the leapfrog method, although it is a symplectic integrator and conserves energy, may introduce numerical precession of the orbit, especially if the timestep is large \citep{springel2005}. In our applications, our choice of a 2nd order N-body integrator should be sufficiently accurate and not change the result substantially. The first reason is the long orbital periods of the binaries: our wide binaries make fewer orbits ($<< 200$ orbits as in \citet{springel2005}) from formation to the end of the simulation. There is a possibility that the orbits may be affected by the type of integrator when there are multiple sink particles. However, this usually happens when a gas disc fragments and this phase lasts only a few kyrs. Hence, we expect that the orbits after the initial fragmentation phase should be insensitive to the integration scheme given the short integration time steps determined by the radiation/chemical time scales and the wide separations of the stars. In addition, the initial phase is highly unpredictable and complex because other factors such as feedback, sink formation criterion, and mergers play a significant role. Therefore, we are confident that our integration method is sufficiently robust and convergent when compared to other uncertainties introduced by the complex physics adopted in the realistic simulation of Pop~III star formation.

\subsection{Torque}
\label{sec:torque}
To study the migration of stars we estimate their orbital angular momentum. The torques are measured in the reference frame where the angular momentum vector is aligned with the $z-$axis. The CoM of the binary is at the origin of the frame and the frame is not accelerated. In Appendix~\ref{app:frame}, we show that the torques on the binary measured in accelerated and non-accelerated frames are equal. The total orbital angular momentum of a binary is,
\begin{equation}
    \label{eq:ang}
    \boldsymbol{J} = \sum_{\subi=1}^{2} \boldsymbol{J}_{\subi} = \sum_{\subi=1}^{2} m_{\subi} \boldsymbol{r}_{\subi} \times \boldsymbol{v}_{\subi},
\end{equation}
where the subscript $i$ indicates the index of a sink in the binary, $\boldsymbol{J}_{\subi}$ is each angular momentum, $m_{\subi}$ is the mass, $\boldsymbol{r}_{\subi}$ is its position from the CoM of the binary, and $\boldsymbol{v}_{\subi}$ is the velocity relative to the CoM. The rate of change of $\boldsymbol{J}$ is,
\begin{equation}
    \label{eq:djdt}
    \frac{\mathrm{d} \boldsymbol{J}}{\mathrm{d}t} = \sum_{\subi=1}^{2} \left( \frac{\mathrm{d}m_{\subi}}{\mathrm{d}t} \boldsymbol{r}_{\subi} \times \boldsymbol{v}_{\subi} + m_{\subi} \frac{\mathrm{d} \boldsymbol{r}_{\subi}}{\mathrm{d}t} \times \boldsymbol{v}_{\subi} + m_{\subi} \boldsymbol{r}_{\subi} \times \frac{\mathrm{d}\boldsymbol{v}_{\subi}}{\mathrm{d}t} \right).
\end{equation}
\begin{figure*}
    \centering
	\includegraphics[width=0.95\textwidth]{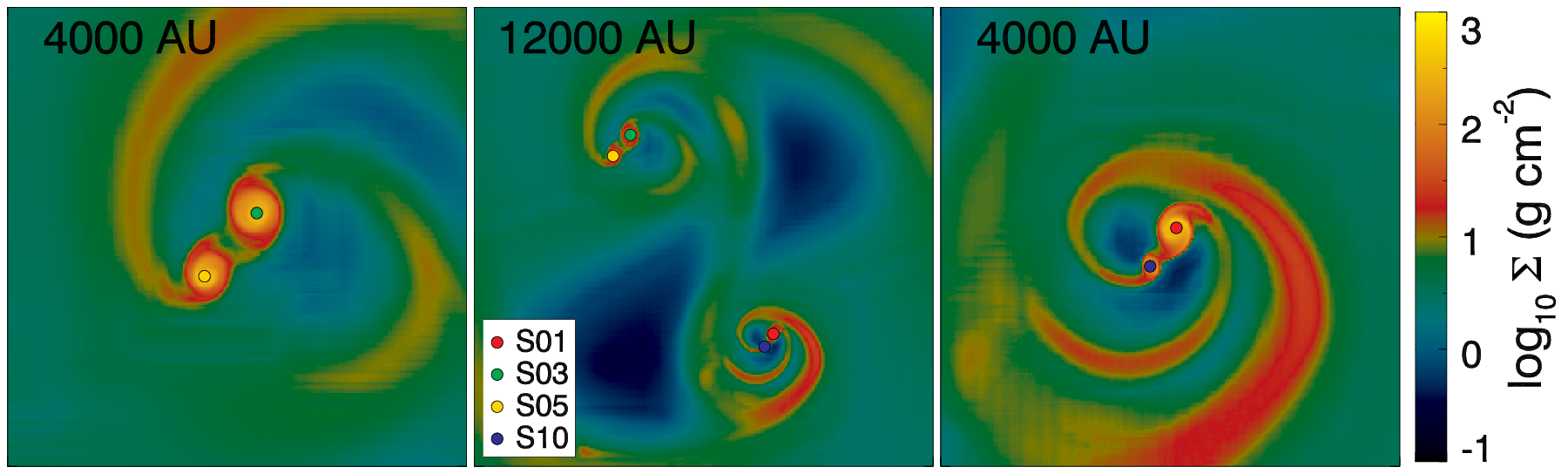}
    \caption{Snapshot of the case \outward. The colour map represents the gas surface density. Sink particles are shown in coloured circles. The hierarchical binary (middle panel) consists of two binaries (left and right panels). The centre of each plot is the CoM of the binary. The FoVs are (4,000~AU)$^2$, (12,000~AU)$^2$, and (4,000~AU)$^2$, respectively.}
    \label{fig:binary}
\end{figure*}

The first term on the right-hand side of the equation describes the change in the angular momentum due to the mass growth of the binary. The change of $J$ caused by the third term is due to external torques exerted on the binary. When gas is accreted smoothly, the second term is zero since $d\boldsymbol{r}_{\subi}/dt \times \boldsymbol{v}_{\subi} = \boldsymbol{v}_{\subi} \times \boldsymbol{v}_{\subi}=0$. This does not hold, however, when a merger with a third body offsets the position of Sink $i$. In this case, the second term is treated as a merger torque (see equation~(\ref{eq:torque_merger})). We can arrange the various terms contributing to the angular momentum change as follows:
\begin{equation}
    \label{eq:total_torque}
    \frac{\mathrm{d} \boldsymbol{J}}{\mathrm{d}t} - \left( \sum_{\subi=1}^{2} \frac{\mathrm{d}m_{\subi}}{\mathrm{d}t} \boldsymbol{r}_{\subi} \times \boldsymbol{v}_{\subi} \right) = \boldsymbol{\tau}_{total} = \boldsymbol{\tau}_{gas} + \boldsymbol{\tau}_{sink} + \boldsymbol{\tau}_{merge} + \boldsymbol{\tau}_{acc}.
\end{equation}
Here, $\boldsymbol{\tau}_{gas}$ is the gravitational torque exerted by the gas,
\begin{equation}
    \begin{split}
        \boldsymbol{\tau}_{gas} &= \sum_{\subi=1}^{2}  \boldsymbol{r}_{\subi} \times \boldsymbol{F}_{gas} \\
        &= \sum_{\subi=1}^{2}  \boldsymbol{r}_{\subi} \times \left( \sum_{\subj=cell} \frac{G m_{\subi} m_{\subj}}{(s^2+\varepsilon^2)^{3/2}} \boldsymbol{s}_{\subj} \right),
    \end{split}
    \label{eq:tau_gas}
\end{equation}
where $\boldsymbol{F}$ is the gravitational force on Sink $i$, $G$ is the gravitational constant, $m_{\subj}$ is the mass of gas cell $j$, $\boldsymbol{s}_{\subj}$ is the displacement vector from the sink to the cell, and $s=\lvert \lvert \boldsymbol{s}_{\subj} \rvert \rvert$ is the distance between them. The gravitational torque exerted by other sinks is calculated similarly,
\begin{equation}
    \begin{split}
        \boldsymbol{\tau}_{sink} &= \sum_{\subi=1}^{2} \boldsymbol{r}_{\subi} \times \boldsymbol{F}_{sink} \\
        &= \sum_{\subi=1}^{2} \boldsymbol{r}_{\subi} \times \left( \sum_{\substack{ {j = \rm sink} \\ j \neq i }} \frac{G m_{\subi} m_{\subj}}{(s^2+\varepsilon^2)^{3/2}} \boldsymbol{s}_{\subj} \right),
    \end{split}
    \label{eq:tau_star}
\end{equation}
where we sum the gravitational torques exerted on Sink $i$ from all the other sinks in the simulation. Note that the torque from the companion star that makes up the binary is zero. For this reason, if a third sink particle does not exist or is far away, this term does not contribute to the evolution of the orbit. $\boldsymbol{\tau}_{merge}$ is the change in angular momentum due to mergers. If a third body merges with Sink $i$, then it has a new position and velocity. The rate of change is,
\begin{equation}
    \boldsymbol{\tau}_{merge} = \frac{1}{\Delta t} \sum_{\subi=1}^{2} m_{\subi,old} \left( \boldsymbol{r}_{\subi,new} \times \boldsymbol{v}_{\subi,new} - \boldsymbol{r}_{\subi,old} \times \boldsymbol{v}_{\subi,old} \right),
    \label{eq:torque_merger}
\end{equation}
where the subscripts "new" and "old" refer to the position and velocity vectors with respect to the CoM of Sink $i$ after and before the merger with a third body, respectively. Finally, we calculate the torque due to the accretion of gas onto the stars 
following \citet{tang2017}.
\begin{equation}
    \boldsymbol{\tau}_{acc} = \sum_{\subi=1}^{2} \sum_{\substack{\rm \textit{j}=cell \\ s \leq r_{sink} \\ \Delta m_{\subj}>0}} \frac{\Delta m_{\subj}}{\Delta t} \boldsymbol{r}_{\subi} \times (\boldsymbol{v}_{\subj} - \boldsymbol{v}_{\subi}).
    \label{eq:tau_acc}
\end{equation}
Note that this term plus $(\mathrm{d}m_{\subi}/\mathrm{d}t) \boldsymbol{r}_{\subi} \times \boldsymbol{v}_{\subi}$ (left-hand side of equation~(\ref{eq:total_torque})) is equal, neglecting numerical errors, to the rate of angular momentum accretion onto the sinks. The latter, however, does not directly contribute to the change in the orbit, and thus only the former is treated as torque in equation~(\ref{eq:tau_acc}). When a gas cell inside the sink radius and its number density exceeds the sink density threshold ($n \geq 10^{12}$~\cc), the linear momentum of the extra mass $\left[\frac{\Delta m_{\subj}}{\Delta t}(\boldsymbol{v}_{\subj}-\boldsymbol{v}_{\subi})\right]$ is dumped to the central sink particle where $\Delta t$ is the time-step. The time evolution of the angular momentum due to each of the torques is calculated by integrating in time each torque,
\begin{equation}
    \label{eq:int_tau}
    \boldsymbol{J}_{\subi} = \int \boldsymbol{\tau}_{\subi} \mathrm{d}t \approx \sum \boldsymbol{\tau}_{\subi} \mathrm{d}t,
\end{equation}
where the subscript $i$ denotes tot, acc, merge, sink, or gas. We also consider the specific angular momentum of the binary,
\begin{equation}
    \label{eq:js}
    \boldsymbol{j} = \sum_{\subi=1}^{2} \frac{\boldsymbol{J}_{\subi}}{m_{\subi}} = \sum_{\subi=1}^{2} ~ \boldsymbol{j}_{\subi} = \sum_{\subi=1}^{2} \boldsymbol{r}_{\subi} \times \boldsymbol{v}_{\subi}.
\end{equation}
Note that $\boldsymbol{J}/M_{b} =(m_1 \boldsymbol{j}_1+m_1 \boldsymbol{j}_2)/M \not= \boldsymbol{j}$, where we have defined the total mass $M_{b}\equiv m_1 + m_2$.

\subsection{Orbital Energy}

In order to quantify how the orbits of stars in a binary respond to torques, we estimate the orbital parameters and their changes following the notation in \citet[][hereafter, M19]{munoz2019}. The orbital energy is
\begin{equation}
    \label{eq:egy}
    \begin{split}
        \mathcal{E}_{b} &= \frac{1}{2}v_{b}^2-\frac{GM_{tot}}{r_{b}} \\
        &= -\frac{GM_{tot}}{2a_{b}},
    \end{split}
\end{equation}
where $r_{b} = \lvert \boldsymbol{r}_{b} \rvert = \lvert \boldsymbol{r}_{2}-\boldsymbol{r}_{1} \rvert$ and $v_{b} = \lvert \boldsymbol{v}_{b} \rvert = \lvert \boldsymbol{v}_{2}-\boldsymbol{v}_{1} \rvert$. $M_{tot}$ is the enclosed total mass including the stars and the gas mass within $r_{2}$ (when $r_{1} \leq r_{2}$). We calculate the orbital energy at every coarse time step of the simulations ($\sim 13$~yr). From the orbital energy and $M_{tot}$ we can get the semimajor axis of the orbit, $a_{b}$. Unlike in M19, the mass of the gas disc is not always negligible in some binaries and therefore we include the enclosed gas mass. Note that the time steps between outputs of hydrodynamics information are limited to $\sim 1-2$~kyr (due to the large size of the files). For this reason, the gas mass between two snapshots is interpolated. Hence, especially for the cases when the gas mass is a non-negligible fraction of the total mass, the estimates of $M_{tot}$ and $\mathcal{E}_{b}$ may become less accurate due to sparse sampling. The orbital eccentricity is,
\begin{equation}
    e_{b} = \sqrt{ 1 + \frac{2l_{b}^2\mathcal{E}_{b}}{(GM_{tot})^2}}.
    \label{eq:ecc}
\end{equation}
In this equation, $l_{b} = \lvert \boldsymbol{l}_{b} \rvert = \lvert \boldsymbol{r}_{b} \times \boldsymbol{v}_{b} \rvert$. This is equal to $J$ divided by the reduced mass (see Equation~(\ref{eq:ang})) but differs from the specific angular momentum defined in Equation~(\ref{eq:js}). The rate of energy change as a function of time is,
\begin{equation}
    \label{eq:egydot}
    \begin{split}
        \frac{\mathrm{d} \mathcal{E}_{b}}{\mathrm{d}t} &= -\frac{G \dot{M}_{tot}}{r_{b}} + \dot{\boldsymbol{r}}_b \cdot \boldsymbol{f}_{ext}.
    \end{split}
\end{equation}
In this work, however, when we plot the rate of orbital energy change we show the left-hand side of the equation, approximating the derivative as it finite time difference ($\Delta \mathcal{E_{b}}/\Delta t$). We leave studying the contribution of different external forces to the orbital energy evolution as future work.

\begin{figure*}
    \centering
	\includegraphics[width=0.95\textwidth]{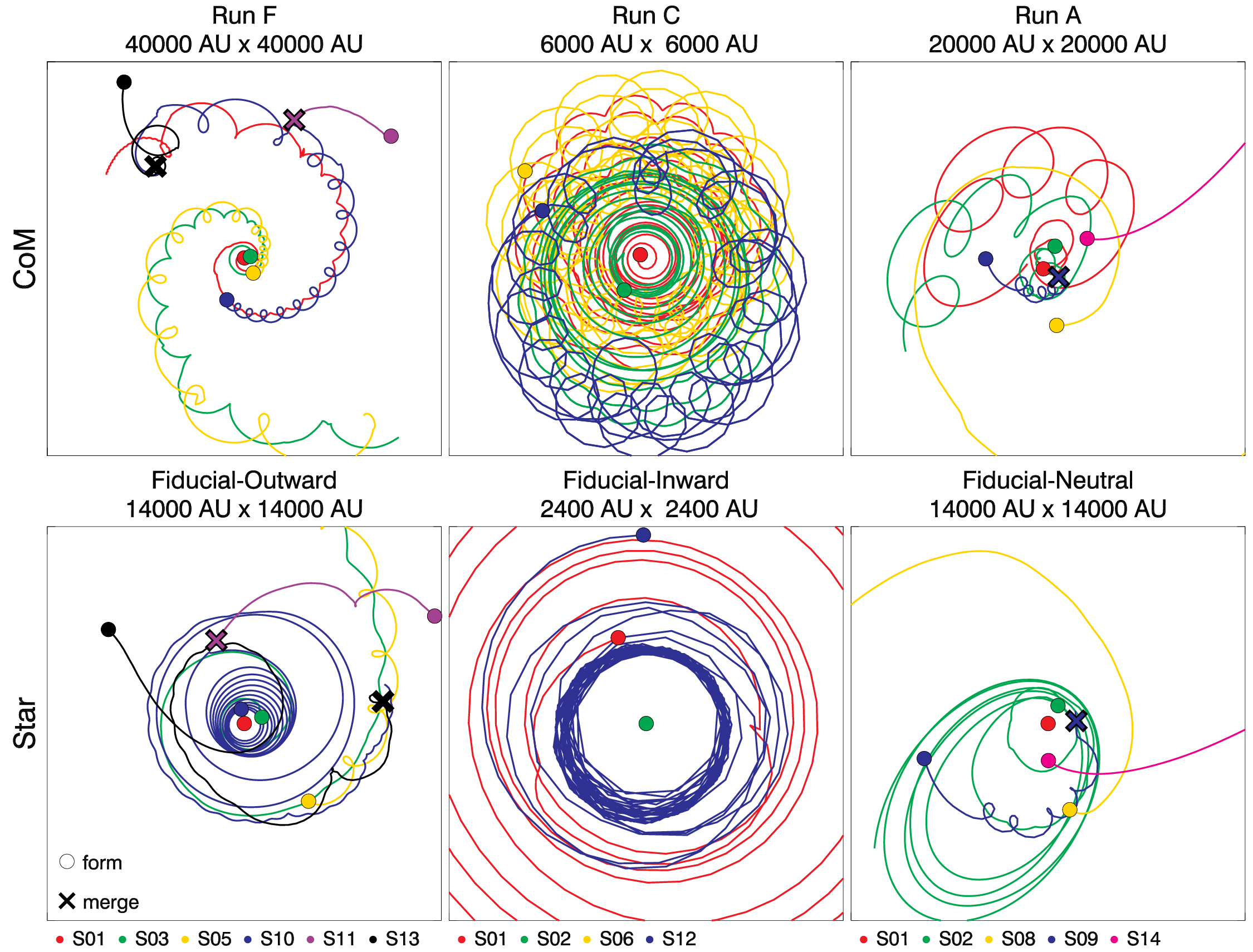}
    \caption{Trajectories of sinks around three representative binaries. Long-lived sinks (more than $10$~kyr) are colour-coded and short-lived ones are not shown here. The FoV of each panel is shown at the top. Circles and crosses indicate the locations of sink formation and mergers, respectively. Top panels: Trajectories in the CoM frame of reference. Bottom panels: Trajectories in the frame of reference centred on one of the binary star. Left panels: Fiducial-Outward binary. S01 (red) is at the centre of the plot. Most long-lived stars migrate outward and have eccentric orbits but an exception is S11 (purple) merging to S10 in Run~F. Middle panels: \inward binary. S02 (green) is at the centre. S01 (red) migrates outwards but S12 (blue) migrates inward. Right panels: \non binary. The orbit of S02 becomes highly eccentric ($e \sim 0.8$) in $2-3$ orbits.}
    \label{fig:orbit}
\end{figure*}


\section{Origin of outward migration}
\label{sec:result1}

\subsection{Typical formation scenario}
\label{sec:general}
In this subsection, we briefly summarise the general scenario of Pop~III star formation in a primordial disc according to the results of our previous papers in the series. For a more detailed picture, see Section 3.1 and Fig.~4 of Paper~III. A gas cloud with a nearly isothermal profile and peak density $\nh \sim 10^7$~\hcc\ contracts and flattens out to form a protostellar disc due to the angular momentum conservation. This gas disc becomes gravitationally unstable and fragments \citep*{kimura2021}, and multiple sink particles form. Some of the sinks migrate inward in a few kyr and merge with others while the others survive and migrate outward. The survivors possess a circumstellar minidisc and these minidiscs may fragment to form multiple stars. Some stars migrate inward while others migrate outward repeating the initial fragmentation time evolution but on the smaller scale of the circumstellar disc (rather than the larger circumbinary disc). In this scenario, the most common outcome is a hierarchical binary (a binary of binaries, Fig.~\ref{fig:binary}), but dynamically unstable systems \citep[e.g., single-triple pair,][]{sugimura2020} may appear. Each of the binaries generally also migrates outward, but there are a few exceptions (Section~\ref{sec:result2}). At late times, in some systems stars can form at Lagrange points L4/L5 of the main binary. This scenario, however, depends on the intensity of the X-ray background that regulates the gas accretion rate onto the disc and stars via enhanced cooling. Strong X-ray irradiation typically lowers the multiplicity and masses of Pop~III stars (see Paper~II and Paper~III).

In Fig.~\ref{fig:orbit} we plot the trajectories of long-lived sink particles (shown as coloured circles at the time of formation and crosses when they merge) for the three runs in which our fiducial binary cases are found. The top panels show the star orbits in a system of reference centred on the CoM, while in the bottom panels the system of reference is centred on one of the binary stars. Sinks that are short-lived are not shown here: they migrate inward on a timescale of a few kyr before merging with other sinks. In general, the orbits of long-lived sinks are eccentric and expand with time. This can be clearly seen in the orbit of the \outward case (red and blue symbols in the bottom left panel). The top left panel also shows the outward migration of the individual binaries. The orbit of this binary expands faster at a later time due to the gravitational torque by S11 (purple) and the merger with it (purple cross). 

In the middle panels, we plot the trajectories of the stars in Run~C (top) and the \inward binary (bottom). With S02 (green) fixed at the centre of the frame of reference (bottom middle panel), its companion (S12, blue) forms at a distance of $\sim 1200$~AU and it migrates inward down to $\sim 400$~AU. Later, the size of the orbit remains nearly constant without further migrating or merging with S02. In the right panels, we show a case (\non: red and green) in which the orbit does not expand significantly but it is highly eccentric ($e \sim 0.8$). In this run at later times S09 (blue) merges with S02 while other stars (S08 and S14) are ejected via three-body interactions.

\subsection{Source of angular momentum of sinks}
\label{sec:outward}
Conservation of angular momentum imposes that, in the absence of external torques and gas accretion from outside the system, the binary orbital parameters remain constant. The migration of the stars, therefore, is caused by the transfer of angular momentum between the binary and other parts of the system via torques and/or gas accretion (\citealp{tang2017}; M19; \citealp{moody2019}; CH19; \citealp{tiede2020,dittmann2022}). The first step to understanding the dominant physical mechanism causing migration, therefore, is to estimate the torques existing in the system. Note that we often loosely refer to the accretion of angular momentum as a torque, $\tau_{acc}$, as defined in equation~(\ref{eq:tau_acc}). 

\begin{figure}
    \centering
	\includegraphics[width=0.48\textwidth]{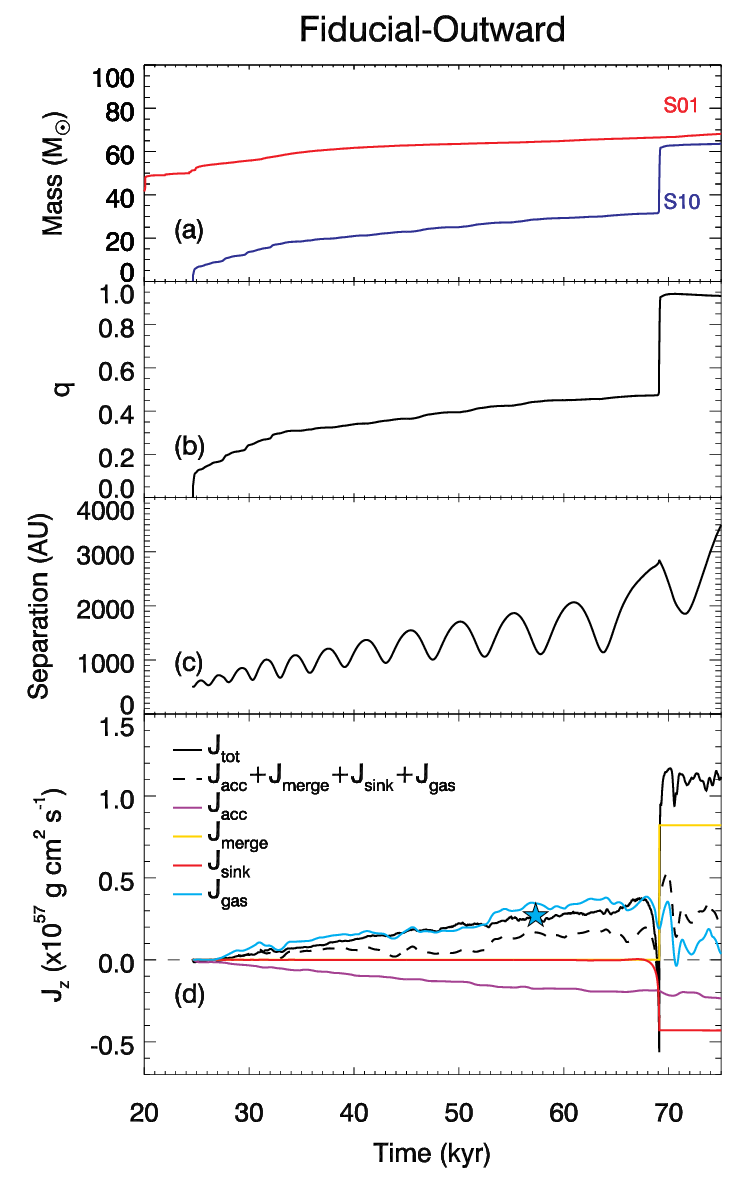}
    \caption{Panel a: The masses of the individual stars in \outward binary. Panel b: The mass ratio between the stars. Panel c: The separation between the stars. Panel d: Predicted and measured (solid line) time evolution of the angular momentum due to the effect of different torques (from equation~(\ref{eq:total_torque}) to (\ref{eq:int_tau})). Only the $z$-components of the vectors are shown in the figure. We distinguish the sum of the angular momentum from individual torques (dashed line) from the actual angular momentum of the binary measured directly (black solid line). As a sanity check, the light blue star indicates the contribution to the angular momentum from the gas disc calculated with the method discussed later and shown in Fig.~\ref{fig:torque_map_sink_total}.}
    \label{fig:torque_out}
\end{figure}

Fig.~\ref{fig:torque_out} shows an example of such calculation for our prototype binary system showing outward migration (\outward). This system is the binary S01-S10 in the minidisc around star S01 in Run F (see also top left panel of Fig.~\ref{fig:orbit}). S10 forms at $t\sim 25$~kyr from the fragmentation of the circumstellar disc around S01 ($M \sim 50~\msun$). It has an initial separation from S01 of $\sim 500$~AU and a mass $q\sim 0.1$ times that of S01. While the binary is accreting gas, the mass ratio $q$ increases and reaches $q \sim 0.5$ before the merger of S10 and S11 (Panel b). This can also be interpreted as a preferential accretion onto the secondary star, also pointed out in previous works \citep{bate2000,farris2014,munoz2020}. In addition, the separation increases by a factor of 4 ($\sim$ 2,000~AU, Panel c). After the merger, this initially unequal mass binary becomes of nearly equal mass (top panel) and the separation increases by a factor of two. Here, however, we focus on understanding the effect of other torques (i.e., accretion and gravitational) before the merger.

\begin{figure}
    \centering
	\includegraphics[width=0.48\textwidth]{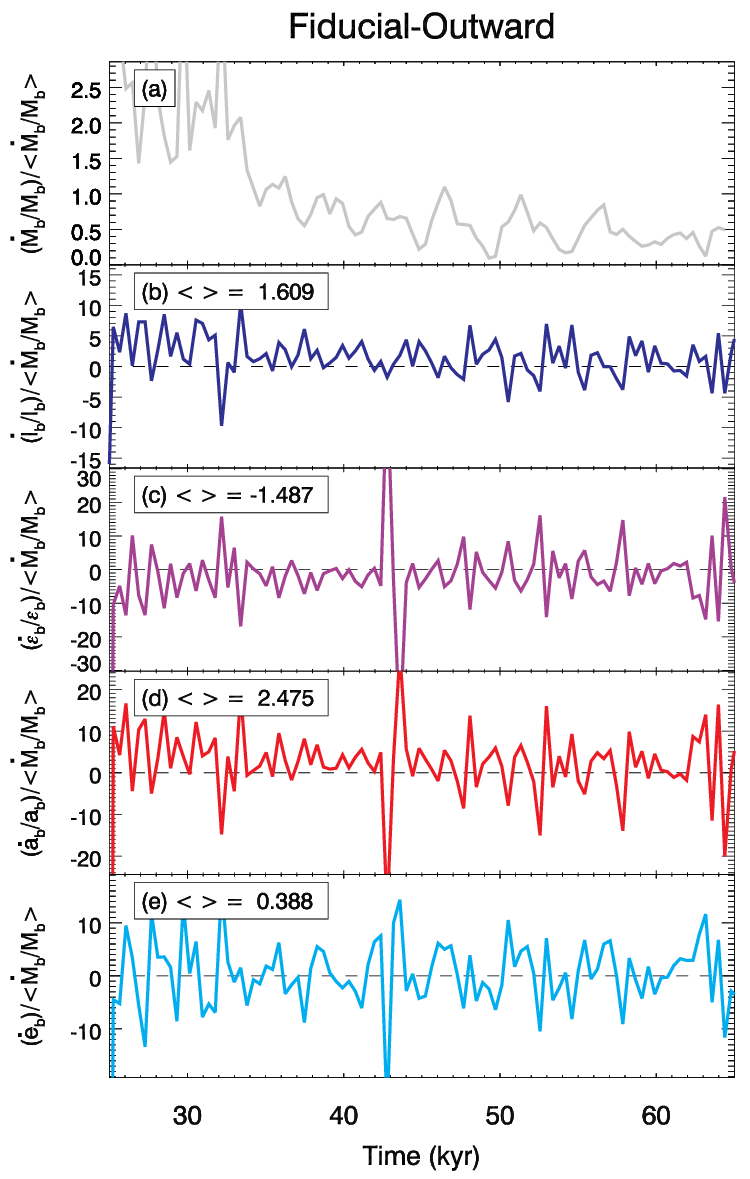}
    \caption{ The evolution of $\dot{M_{b}}/M_{b}$ (Panel a), $\dot{l_{b}}/l_{b}$ (Panel b), $\dot{\mathcal{E}_{b}}/\mathcal{E}_{b}$ (Panel c), $\dot{a_{b}}/a_{b}$ (Panel d), and $\dot{e_{b}}$ (Panel e) in the \outward binary. The rates in Panel~a to e are normalised by $<\dot{M_{b}}/M_{b}>$ (Panel~a). We show the first $\sim 40$~kyr before a merger. At the top left corner of each panel, we show the average value (except Panel~a, the average is unity). Each plot is smoothed over 30 points to show the general trend better. In Panel~b and c, we found that external forces lead to an increase in angular momentum and orbital energy. The orbit expands as shown in Panel~d. $<\dot{e}_{b}>$ is positive meaning the orbit becomes more eccentric (Panel~e). }
    \label{fig:energy_out}
\end{figure}
\begin{figure*}
    \centering
	\includegraphics[width=0.95\textwidth]{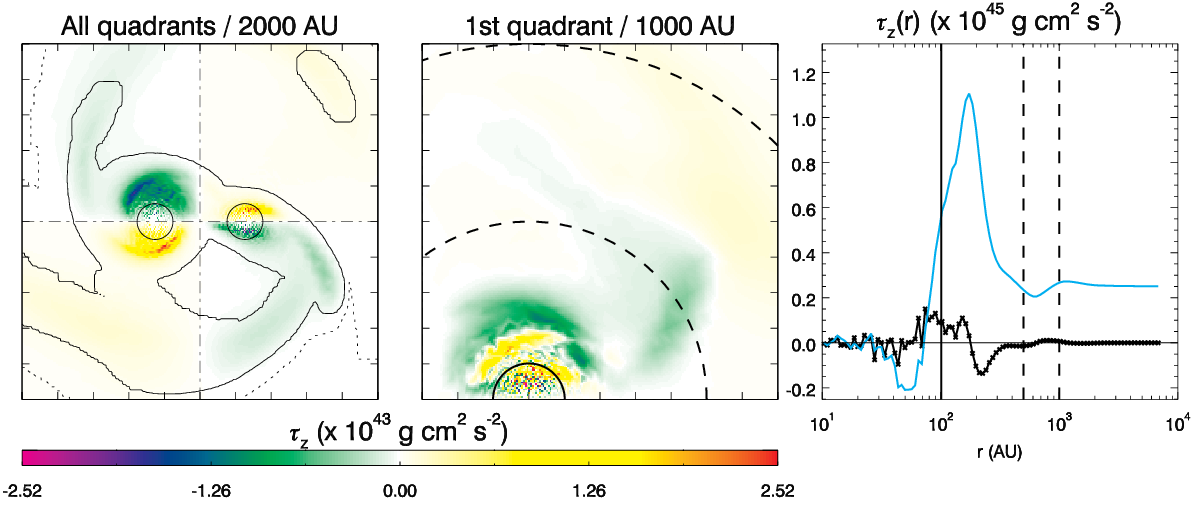}
    \caption{Left: Gravitational torque map right after the formation of the binary in the  \outward binary ($t \sim 25$~kyr). We calculate the gravitational torque exerted on the binary from gas cells. The torques are projected onto the logarithmic polar grid ($r$ and $\theta$) around each star. The angular momentum vector of the binary is perpendicular to each plane. The density on the left half of the figure is projected onto the polar grid around S01 (left circle) and the density on the other side is projected onto the grid around S10 (right circle). In addition, we cover the gas cells with height $\lesssim 41000$~AU along the line-of-sight and, therefore no torque is effectively missing. Circles are sink particles with radii to scale. The centre of the plot is the midpoint between two stars. The midpoint is chosen as the centre so that two sinks are overlapped when the map is folded (see the description of the middle panel). The iso-density contours indicate $\nh = 10^9$~cm$^{-3}$ and $10^{10}$~cm$^{-3}$ (dotted and solid lines, respectively). The binary is orbiting in the counterclockwise direction. The FoV is 2,000~AU $\times$ 2,000~AU. Middle: Net torques folded and added to the first quadrant. The torque map in the left panel is folded twice along the dotted-dashed lines so that the two sinks overlap. The FoV is 1,000~AU $\times$ 1,000~AU. Dashed lines are two concentric circles with radii 500 and 1,000~AU. Right: The torque as a function of distance. We measure the torque in the middle panel and $r$ indicates the distance from the sink particles (centre of the semicircle) in the disc plane. We do not show the torques with $r<10$~AU because this scale is smaller than the spatial resolution of the simulation ($12.6$~AU). The black line is the torque at r and the blue line is the cumulative torque. The vertical lines show the locations of the sink radius (solid line) and 500~AU and 1,000~AU (dashed lines).}
    \label{fig:torque_map_sink}
\end{figure*}
\begin{figure}
    \centering
	\includegraphics[width=0.48\textwidth]{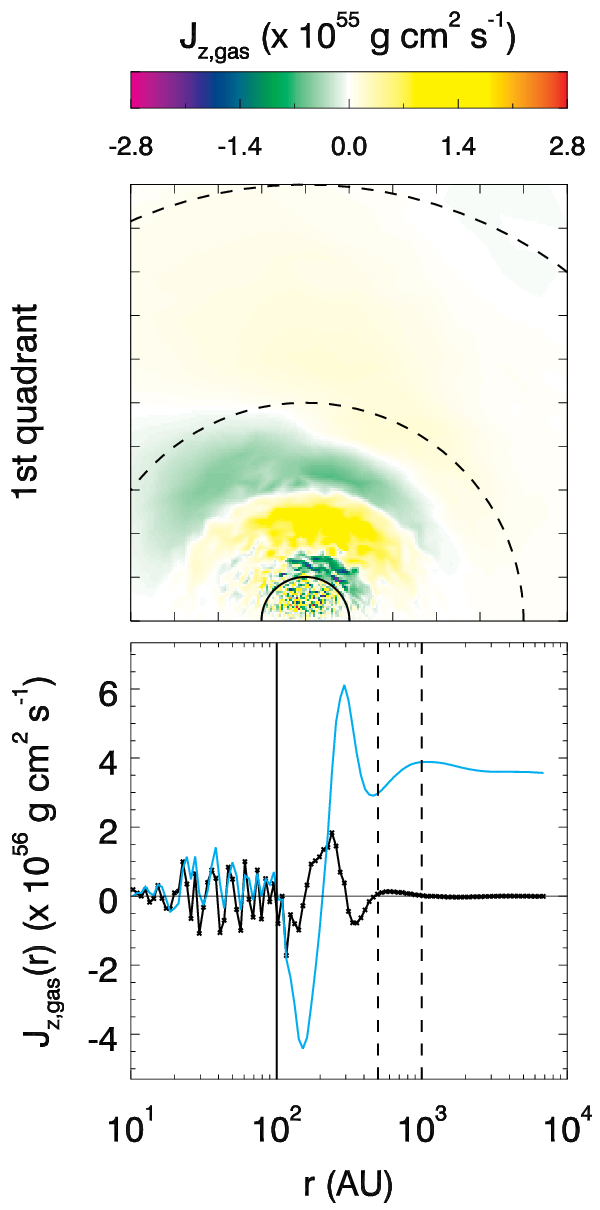}
    \caption{Top: Folded torque map (see caption for the middle panel of Fig.~\ref{fig:torque_map_sink}) integrated over time. Therefore, the colour map represents the contribution to the change in angular momentum of the binary from different regions of the gas disc. The sinks are dragged in the direction of the motion by the gas in front of them (yellow region). Bottom: Radial profile of torque. The format of this panel is the same as that in the right panel of Fig.~\ref{fig:torque_map_sink}.  The figure shows the values at $t \sim 57$~kyr. As a sanity check, we plot the total torque in Panel d of Fig.~\ref{fig:torque_out} with a light blue star. }
    \label{fig:torque_map_sink_total}
\end{figure}
Before exploring the main mechanism for the outward migration, we try to understand the outward migration by looking at the changes in the parameters related to the binary. Following M19, we plot the evolution of $\dot{M_{b}}/M_{b}$, $\dot{l_{b}}/l_{b}$, $\dot{\mathcal{E}_{b}}/\mathcal{E}_{b}$, $\dot{a_{b}}/a_{b}$, and $\dot{e_{b}}$ for the \outward binary in Panel~a to e of Fig.~\ref{fig:energy_out}. As the original plot is noisy, we smoothed each plot by averaging the values for every $\Delta t \approx 0.4$~kyr (30 data points). In M19, these values are shown as a fraction of the prescribed constant gas inflow rate ($\dot{M}_0/M_{b}$) but this quantity is unavailable in our non-idealised simulation. For this reason, we normalise those rates with the average accretion rate $<\dot{M_{b}}/M_{b}> \approx 1/t_{acc}$ in Panel~a. The accretion time scale $t_{acc}$ of \outward is $\sim 81$~kyr. The net force exerts a torque on the binary and delivers energy to it (Panel~b and c) causing the binary to migrate outward (Panel~d). The initial eccentricity of the binary is 0.14 and increases with time (Panel~e). It is unlikely, however, that the gas disc drives this evolution toward the eccentricity attractor $\sim 0.4$ (\citealp{zrake2021}; M19; \citealp{dorazio2021,lai2023}) because other sink particles and mergers appear to be the main factors responsible for a substantial increase of the eccentricity (see discussed in Section~\ref{sec:ecc}).

To understand the main mechanism for the outward migration we compare the increase of angular momentum by separating the effect of various torques as a function of time (see bottom panel). Before the merger, the gravitational torque from the disc (light blue line) and accretion of angular momentum (purple line) are the two dominant sources of angular momentum (see bottom panel). However, they work in opposite directions: the gas accretion torque is negative, hence reducing the binary angular momentum and acting as a viscous or drag term. The gravitational torque is instead positive (producing outward migration) and it dominates over the negative torque. The net change in $J_z$ from the sum of all torques (dashed line) is positive. When compared to the actual orbital angular momentum of the binary (black solid line) the agreement is not perfect because the calculations are somewhat uncertain due to the finite time steps of our outputs ($\sim 1-2$~kyr). The result shown in Fig.~\ref{fig:torque_out} is representative of most simulations in that the gravitational torque is the dominant term and the accretion torques have a negative sign, somewhat reducing the outward migration effect. When stars are observed to migrate outward, both the solid and dashed lines are increasing nearly in all binaries we analysed, meaning the gas disc plays the dominant role in producing the binary migration via gravitational torque.

We established that the outward migration of stars is caused by the gravitational torque by the gas disc. Now we need to understand why the sign is positive (i.e., the star is accelerated in the direction of its rotation by some gas overdensity in front of it) and which part of the gas disc contributes most to the torque (outer disc, the spiral structure or the minidisc). There is a general agreement that the outer gas disc exerts negative torque on the binary (\citealp{gould2000,armitage2002,chang2010}; M19; \citealp{dittmann2022}) thereby shrinking it. In the context of Pop~III binaries, CH19 also demonstrated that disc spiral structures extract the angular momentum from the binaries. Hence, the question becomes whether other components in the system provide positive torque to overcome the effect of the outer disc. \citet{tang2017} argued that the gravitational torque by the minidiscs around sink particles (SMBHs in this case) drives inward migration. On the other hand, M19 demonstrated the torque from the inner circumstellar discs is dominant over the outer torques and is responsible for the outward migration. In our simulations, as observed in CH19, a spiral structure often appears when newly formed sinks migrate inward to merge with the central one. Once this initial merger phase is over, the remaining stars have circumstellar minidiscs and spiral arms extending outward and connecting the minidiscs (see Fig~\ref{fig:binary}). For this reason, we hypothesise these non-axis-symmetric structures are the dominant contribution to the torque and outward migration of the binary. To test this hypothesis, we construct maps showing the contribution to the gravitational torque on the binary from each gas cell on the disc (shown face-on). To visualise these maps we choose a 2-dimensional logarithmic polar grid ($r$ and $\theta$) around each sink particle and calculate the total torque on the binary contributed by each cell on this grid. These grids are on the orbital plane of the binary and are perpendicular to the binary angular momentum vector. This approach imposes the cell size $r^2 \mathrm{d}\theta \mathrm{d}\ln{r}$ and thus eliminates $r^{-2}$ dependence of gravity which might lead to visually underestimating the effect of outer spiral arms despite their large extent and mass. The parameters of the logarithmic grid are provided in Table~\ref{tab:polar}. Note that the colour maps show the total torque in each cell, not torque density like in previous works (\citealp[e.g.,][]{tang2017}; M19; \citealp{dittmann2021,dittmann2022}): in this work the torque density would also lead to visually underestimating the role of the outer structures because of the increasing cell size with radial distance. The contribution to the torque from different parts of the gas disc is visualised in the left panel of Fig.~\ref{fig:torque_map_sink} for \outward binary. Although the torque by the minidiscs is large, spiral arms also exert significant torques on the binary, although the sign of the torque varies between positive and negative values making it difficult to determine if the net torque is positive or negative. For quantitative analysis, we fold the torque map twice along the $x$-axis and vertical line crossing the midpoint of two sinks (dotted-dashed lines). This approach is novel and we found only one other study folding torque maps similarly to here \citep{li2022}. Unlike in \citet{li2022}, however, our binaries are unequal mass, hence the torque maps on each side are not symmetric. We highlight, however, that our approach distinguishes the torque by the two circumstellar minidiscs from that by the outer gas disc and shows that they play a major role in producing outward migration (as discussed in the next paragraph). In this way, torques in the entire domain are added to the 1st quadrant and it is easier to determine the relative importance of the positive and negative regions of the torque and their location. As can be seen in the middle panel, the torque is positive near the sink radius (solid line) and outside ($\sim 200$~pc, top left of the solid line). This region corresponds to the front side of the minidiscs and therefore this result implies that the sinks are accelerated in the directions of the binary orbital velocity. The azimuthally averaged radial profile, shown in the right panel, also shows that the gas just outside the sinks dominates the contribution to the net positive torques and therefore is the cause of the outward migration.

\begin{table}
    \caption{Parameters of the logarithmic polar grid.}
    \footnotesize
    \begin{threeparttable}
        \centering
        \label{tab:polar}
        \begin{tabular}{ | l | l | l | }
		\hline
            Parameter & Value & Description \\
		\hline
            $r_{min}$ & $10$~AU & inner boundary \\
            \hline
            $r_{max}$ & $7071$~AU & outer boundary \\
            \hline
            $N_{r}$ & $100$ & number of grids in the radial direction \\
            \hline
            $N_{\theta}$ & $60$ & number of grids in the azimuthal direction \\
            \hline
            ${\rm d}\ln{r}$ & $0.0656$ & radial resolution \\
            \hline
            ${\rm d}\theta$ & $0.105\pi$ & azimuthal resolution\\
            \hline

	\end{tabular}
    \end{threeparttable}
\end{table}

Fig.~\ref{fig:torque_map_sink} illustrates the importance of the minidiscs in driving the outward migration of the binary at a specific time. However, it does not show the overall effect integrated over time. As seen in Fig.~\ref{fig:orbit} and Fig.~\ref{fig:torque_out} the stars in the binary migrate continuously. This means that the net positive torque has to be kept during this period. This is visualised in Fig.~\ref{fig:torque_map_sink_total}, showing the same torque map integrated over time (i.e., showing the angular momentum, equation~(\ref{eq:int_tau})). In the figure, the bright yellow feature (top panel) and the cumulative torque (blue line, bottom panel) indicate the gas near the sink (minidiscs) exert net positive torque consistently. This result, therefore, implies the minidiscs play the most significant roles in producing the migration of the stars as found in \citet{tang2017}, although contrary to the BH binary case in their work, in our case the sign of the torque is positive and the migration is outward.

\subsection{Source of angular momentum of disc}
\label{sec:disc_out}
As mentioned in the previous subsection, the minidiscs are the dominant source of the gravitational torque on the stars. This means that they inevitably lose some of their angular momentum by dragging their central stars with them. For the circumstellar minidiscs and their central stars to keep migrating outward, therefore, there must be an external source of angular momentum acting on the minidiscs. An obvious candidate is the accretion of high angular momentum gas from the circumbinary envelope. This idea was first suggested in \citet{sugimura2020} and in Paper~II, without a quantitative analysis of this effect. However, Paper~II provided several hints that outward migration is connected to the gas accretion rate. The global rate of gas accretion from larger scales to the disc center ($\dot{M}$) is proportional to $c_{s}^3$, where $c_s$ is the gas sound speed. Even though it seems counter-intuitive, the sound speed is reduced if the gas is irradiated by a strong/moderate X-ray background because of the enhanced formation and cooling by \hm. The reduction in $\dot{M}$ slows down the accretion of high angular momentum gas onto the circumstellar discs and stars. We therefore found that, when irradiated by an X-ray background, Pop~III protostars tend to have smaller masses and the rate of outward migration is reduced, producing binaries with smaller separations (Fig.~8 of Paper~II). We concluded that the rate of gas accretion plays an important role in the angular momentum supply and outward migration of Pop~III binary stars.

\begin{figure}
    \centering
	\includegraphics[width=0.48\textwidth]{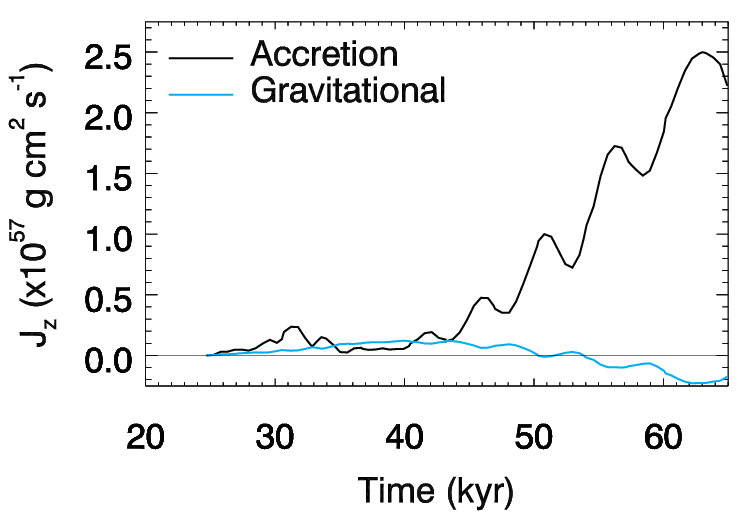}
    \caption{Angular momentum of the minidiscs as a function of time for the \outward binary due to the gas accretion torque (black line) and external gravitational torque (cyan line). Only the z-component of the angular momentum is considered.}
    \label{fig:accJ}
\end{figure}

\begin{figure}
    \centering
	\includegraphics[width=0.48\textwidth]{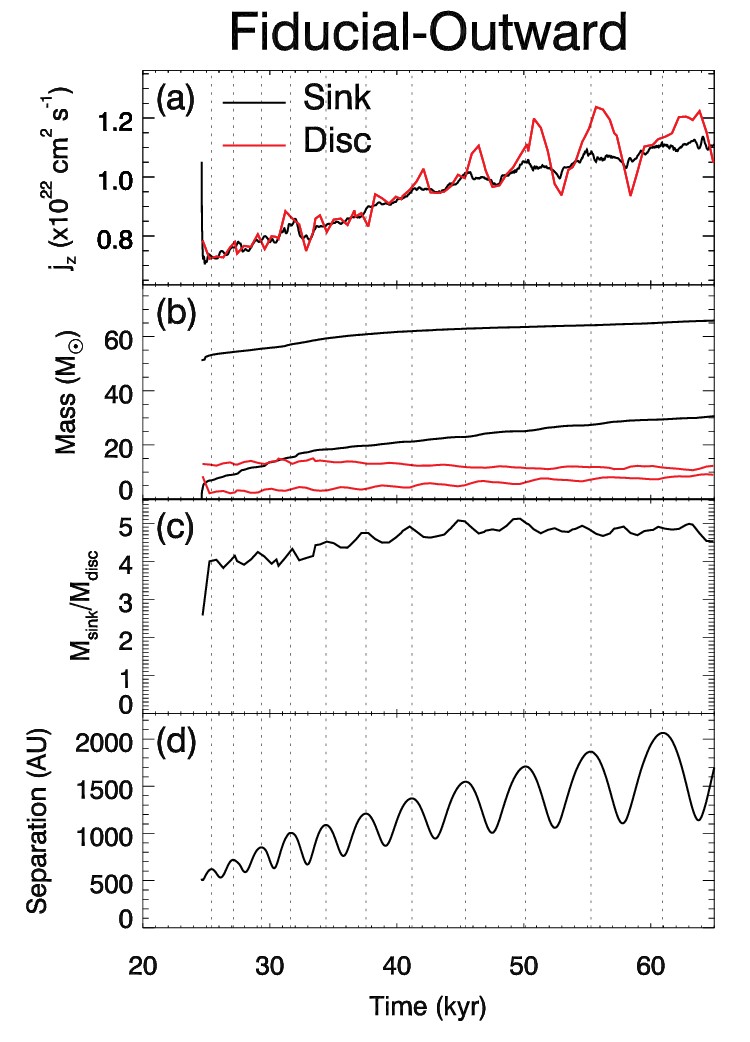}
    \caption{Properties of the \outward binary. Vertical dashed lines indicate maximum separations between two stars. Black lines in Panels a and b refer to stars (sinks) properties while the red lines refer to the same quantities for the gas discs. The discs are defined as gas within a cylinder with a diameter of 600~AU and height of 600~AU. Panel a: Specific angular momentum of the stars and minidiscs (equation~(\ref{eq:js})). Overall, both the stars and discs angular momenta increase with time but that of the discs shows larger fluctuations as a function of time with peaks near the apocentres. Panel b: Time evolution of the masses of individual stars and discs. Panel c: Ratio of the mass in sinks to the mass in the minidiscs as a function of time. The system mass is dominated by the stars and therefore the disc rotation outsided of the stars' outer orbit is nearly Keplerian. The oscillation of the ratio is due to the mass increase of the discs at apocentres. Panel d: Separation between the two stars as a function of time.}
    \label{fig:disc_sink}
\end{figure}

To test the idea that minidiscs gain angular momentum by accreting high angular momentum gas from the outer parts of the disc, we estimate the external gas gravitational torque and angular momentum accretion rate of the discs. To carry out this calculation, we first define a cylinder with a radius $r_{cyl} = 300$~AU and height of $h_{cyl} = 600$~AU around each sink particle. We assume that this cylindrical boundary contains the minidiscs around each sink particle. The gravitational torque is calculated as explained in Section~\ref{sec:outward} for the case of the sinks, using equation~(\ref{eq:tau_gas}). In the equation, however, the subscript $i$ runs over all gas cells within the cylinder and $j$ runs over gas cells outside the cylinder. The change in angular momentum over time due to this effect is shown in light blue in Fig.~\ref{fig:accJ}. To calculate the accretion rate of angular momentum, we identify the cells near the cylinder walls ($r_{cyl}-\Delta x_{min}/2 \leq r < r_{cyl}+\Delta x_{min}/2$), where $r$ is the radial distance in the cylindrical coordinates. The radius of the cylinders is kept fixed for simplicity. The total accretion rate through the walls is calculated as follows:
\begin{equation}
    \frac{ \mathrm{d}J_{\textit{z}} }{\mathrm{d}t} = \sum_{\subi=1}^{2} \sum_{\subj=cell} \frac{J_{\subj,\textit{z}}}{V_{\subj}} v_{\subj,\textit{r}} \Delta x_{\subj}^2.
\end{equation}
The subscripts $i$ and $j$ indicate the sinks and gas cells, respectively. $V_{\subj}$ is the volume of cell $j$ and therefore $(J_{\subj,\textit{z}}/V)$ is its angular momentum density. $v_{\subj,\textit{r}}$ is the radial velocity in the cylindrical coordinate. As seen in Fig.~\ref{fig:accJ}, the change in angular momentum due to gas accretion (black) is dominant over the gravitational torque (light blue) on the disc. Other binaries with outward migration follow a similar trend with a more pronounced difference. Another result that can be noticed inspecting the accretion of angular momentum as a function of time in Fig.~\ref{fig:accJ}, is that the angular momentum accretion has periodic oscillations with peaks happening just after the stars reach the apocentre of the elliptical orbit. This aspect is more clearly seen in Fig.~\ref{fig:disc_sink}, showing that the disc angular momentum (red lines in panel a) peaks near the maximum binary separation (vertical dotted lines). This is because stars are further from the CoM and therefore they can accrete high angular momentum gas more easily. These accretion peaks are pronounced when the secondary star passes the outer spiral arm. At pericentres, on the other hand, the discs lose angular momentum to the sinks, but the net effect integrated over time is toward an increase of the angular momentum.

\begin{figure}
    \centering
	\includegraphics[width=0.48\textwidth]{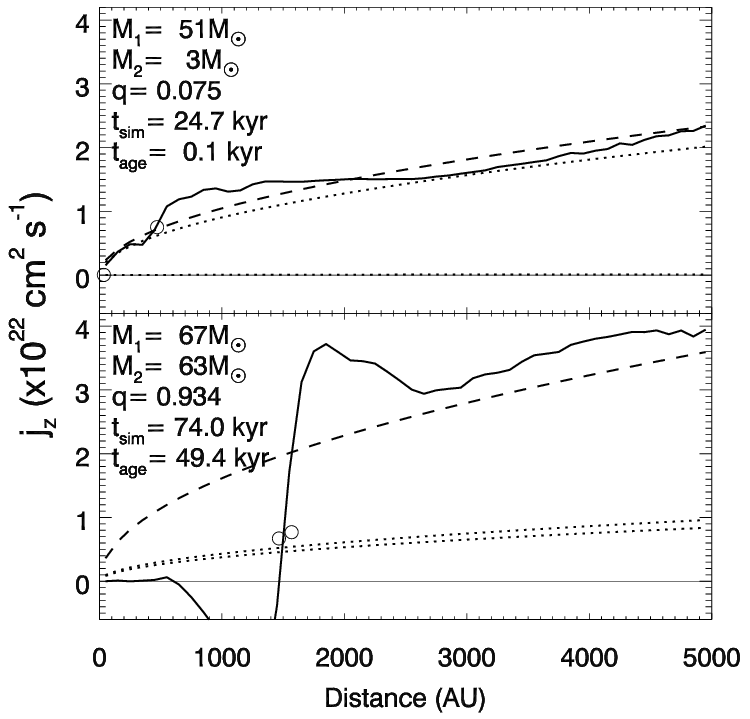}
    \caption{Specific angular momentum of various components in the \outward binary as a function of distance from the CoM (i.e., the centre of the frame of reference). The solid line shows the specific angular momentum of the gas measured at distance $r$. Circles indicate the position and specific angular momenta of the stars. The two dotted lines show the Keplerian specific angular momentum of the individual stars ($j_{star,1}, j_{star,2}$), and the dashed line the Keplerian specific angular momentum $j_{gas,kep}$ (see equation~(\ref{eq:gaskep})). The top and bottom panels refer to two different epochs in binary evolution. The individual masses, the mass ratio ($q$), simulation time ($t_{sim}$), and binary age ($t_{age}$) are shown in the legend. Top panel: Binary soon after its formation ($t_{age} \sim 0.1$~kyr). Note that $j$ of the primary is an order of magnitude smaller than that of the secondary. At large radii, the gas angular momentum is close to that of the secondary star ($M_2$) although it deviates near the secondary due to the presence of a minidisc. Bottom panel: Binary at $t_{age} \sim 39$~kyr since its formation. Since the two stars have comparable masses, their velocities and thus specific angular momenta are significantly lower than the measured gas specific angular momentum at their distance (solid line) that is approximated by the Keplerian specific angular momentum dominated by the star masses (dashed line).}
    \label{fig:J}
\end{figure}

\subsection{Efficiency of angular momentum accretion}
\label{sec:eff}
For binaries to gain angular momentum and migrate outward by gas accretion, the specific angular momentum of the accreted gas must be higher than that of the stars. The difference between these two values, in addition to the mass accretion rate, may explain why some stars migrate outward efficiently while others do not. The answer to this question requires an understanding of the distribution of the angular momentum of gas and stars. 
The specific angular momentum of the stars in a binary is,
\begin{equation}
    \begin{split}
        \label{eq:jstar}
        j_{star} &= j_{star,1} + j_{star,2}=r_1^2 \omega_{kep} + r_2^2 \omega_{kep}  \\
        &= r_1^2 \sqrt{GM/r^3} + r_2^2\sqrt{GM/r^3} \\
        &= \left(\frac{m_2}{M}\right)^2 r^2\omega_{kep} + \left(\frac{m_1}{M}\right)^2 r^2\omega_{kep},
    \end{split}
\end{equation}
where $\omega_{kep} =2\pi/P=\sqrt{GM/r^3}$ is the angular velocity of the binary, with $r=r_1+r_2$, $r_1/r=m_2/M$ and $r_2/r=m_1/M$. Without loss of generality, we assume that $m_2 \le m_1$ (or $q\equiv m_2/m_1 \le 1$), hence the orbit of (Star 2) is further from the CoM of the system. The radial profile of the specific angular momentum of the disc beyond the secondary minidisc is approximately Keplerian as seen in Fig.~\ref{fig:J} (solid and dashed lines). The discs are $\sim 4-5$ times less massive than the Pop~III stars (Panel~c of Fig.~\ref{fig:disc_sink}), which is a non-negligible fraction in terms of mass. Due to its $\sqrt{M}$~dependence, however, this translates into a $\sim 10$~\%~deviation of the rotational velocity with respect to perfect Keplerian case. The profile inside the orbit deviates from Keplerian. However, this is irrelevant to the analysis because we focus on the accretion of angular momentum from the outer disc. The profile near each sink particle also deviates from the Keplerian value because of the gas motion around the sink. This deviation is especially evident in the bottom panel of Fig.~\ref{fig:J} (between $\sim 500$ and $\sim 2000$~AU). However, the main takeaway of this plot is that the binary's angular momentum (dotted lines and circles) is significantly lower than the Keplerian rotation value of the gas disc (dashed line) and thus the binary (or more precisely the minidiscs) gain angular momentum from accreting gas from the outer disc. We approximate the specific angular momentum of the gas beyond the outer star orbit (i.e., at a distance $r_2$ from the CoM) to the Keplerian value as follows,
\begin{figure}
    \centering
	\includegraphics[width=0.48\textwidth]{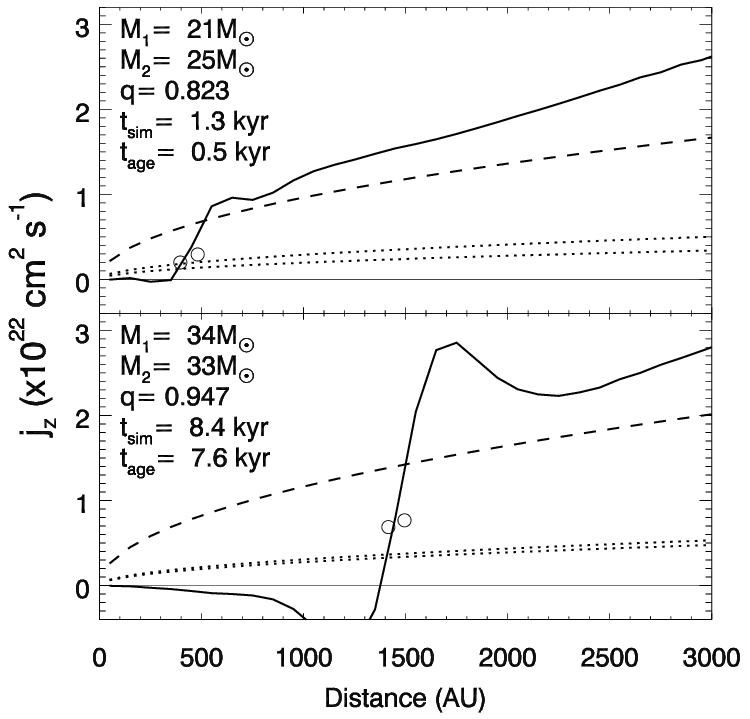}
    \caption{Same as Fig.~\ref{fig:J} but showing the specific angular momentum of the S01-S03 binary in Run~F right after the formation of S03 ($t = 1.3$~kyr).}
    \label{fig:J_init}
\end{figure}
\begin{equation}
    j_{gas, kep}(r_2) \approx r_2 v_{kep}(r_2) = \sqrt{GMr_2},
    \label{eq:gaskep}
\end{equation}
where $v_{kep} = \sqrt{GM/r_2}$ is the Keplerian velocity. If we compare the specific angular momentum of the gas at the same distance of the outer star orbit ($r_2$) we have:
\begin{equation}
    j_{gas, kep}(r_2)/j_{star,2} \approx (r/r_2)^{3/2} = (M/m_1)^{3/2} = (1+q)^{3/2}\ge 1.
\end{equation}
This ratio approaches unity when the mass ratio of the binary is $q=m_2/m_1 \ll 1$ and reaches a maximum of $2^{3/2}$ for equal mass binaries (i.e., $q=1$). In this latter case, both stars in the binary contribute equally to the accretion of higher angular momentum gas. We point out that \citet{sugimura2023} has interpreted the cause of outward migration similarly. We assume the angular momentum vectors are perpendicular to the orbital plane and thus the quantities in the equations above are scalars. The actual distribution of angular momentum is shown in Fig.~\ref{fig:J}. The solid line, two dotted lines, and a dashed line in each panel indicate $j_{gas}$ (measured), $j_{star,1~or~2}$, and $j_{gas, kep}=\sqrt{GMr}$, respectively. As can be seen in the figure, the actual measured values for the sinks (circles) agree with having Keplerian orbits (dotted lines). The figure illustrates how the specific angular momentum distribution varies with the mass ratio. When $q$ is small (top panel), the difference between $j_{gas}$ (solid line) and $j_{star,2}$ (dotted line) is relatively small. When the two stars have comparable masses (bottom panel), on the other hand, individual stars have smaller angular momentum (dotted lines below dashed and solid lines) and thus have a greater difference. Consequently, a binary obtains angular momentum more easily in an equal-mass binary.

The gas motion deviates from the perfect Keplerian motion due to the presence of gas. This deviation is greatest right after the initial fragmentation and the formation of the first binary as shown in Fig.~\ref{fig:J_init} (Run~F at $t=1.3$~kyr). Since a larger amount of gas is available in the beginning, the gas angular momentum scales linearly rather than with $\sqrt{r}$ (Keplerian). Because the infalling gas has a higher specific angular momentum than the sinks, the stars that survive mergers migrate outward efficiently. 

\begin{figure}
    \centering
	\includegraphics[width=0.48\textwidth]{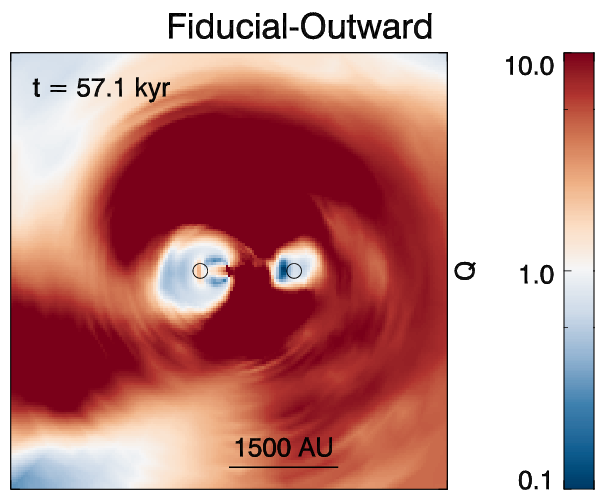}
    \caption{Snapshot of the toomre-$Q$ parameter for the \outward binary at $t = 57.1$~kyr. The regions with low $Q$ around the sink particles (circles) indicate the circumstellar discs are toomre unstable and hence subject to fragmentation The circumbinary disc has $Q>1$, hence it is gravitationally stable at this time during its evolution.}
    \label{fig:Qout}
\end{figure}

This mechanism also explains why the mass ratio of the binaries often approaches $q=1$ for several of the binaries in this work (see Panel b of Fig.~\ref{fig:torque_out}) and in previously published works \citep{bate2000,munoz2020}.  When $q$ is low, the outer star intercepts most of the gas from the circumbinary disc and it grows faster than the primary star. By doing so, $q$ approaches unity and the binary can migrate outward more rapidly.

Finally, we measured the Toomre-$Q$ parameter to figure out the main driving mechanism of the gas accretion. In this subsection, we explain the efficiency of the accretion assuming the gas disc is Keplerian. If the gas disc is gravitationally unstable, the instability-driven spiral arms may exert torques and affect the angular momentum transfer within the Keplerian disc. In Fig.~\ref{fig:Qout}, we plot $Q$ of the gas disc for the Fiducial-Outward binary at $t=57.1$~kyr. We find that at this time the CBD around the binary is Toomre-stable ($Q > 1$), unlike the marginally unable CBD found by \citet{sugimura2023}. Therefore, we speculate spiral arms are not produced by fragmentation/gravitational instability in this phase. The difference seen in the circumbinary disc stability in this work and in \citet{sugimura2023} is beyond the scope of this paper, so we leave it as future work.

To summarise, in this section, we explore the origin of the outward migration of binary stars. The outward migration takes place in a two-step process. First, the minidiscs around stars gain angular momentum by accreting high angular momentum gas. Then the stars embedded in the circumstellar minidiscs gain angular momentum from them via gravitational torque and migrate outward. 

\begin{figure}
    \centering
	\includegraphics[width=0.48\textwidth]{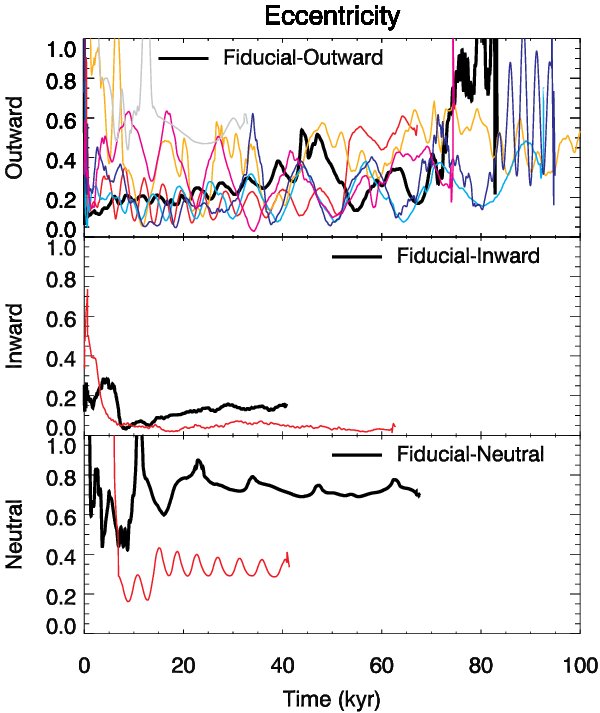}
    \caption{ Time evolution of the orbital eccentricities (equation~\ref{eq:ecc}) for the three migration groups (from top to bottom). All the simulations are colour-coded and the fiducial systems are highlighted with thick solid lines. The $x$-axis indicates the time since the formation of the binary.}
    \label{fig:ecc}
\end{figure}

\subsection{Origin of eccentric orbits}
\label{sec:ecc}
Many Pop~III binaries have eccentric orbits. This is interesting because the eccentricity often gives rise to a periodic variability of Pop~III stars' luminosity and spectral type as discussed in Paper~III. In addition, the existence of highly eccentric orbits (e.g., \non with $e=0.8$) suggests the possible relevance of a new channel for GW emission: binary black hole mergers via dynamical excitation \citep{michaely2019,michaely2020}. These reasons motivate us to further explore the origin of eccentricity. Fig.~\ref{fig:ecc} shows the orbital eccentricity since the formation of each binary in the three migration groups (from top to bottom). We highlight the fiducial cases with thick solid lines. The binaries are generally born eccentric through 3-body or N-body interactions/mergers with other sink particles or gravitational torque by non-axisymmetric structures. In general, the eccentricity oscillates due to the influence of the gas disc and other sink particles. For the cases with outward migration or no migration, the eccentricities oscillate around a nearly constant value (between $e \sim 0.2$ and $0.4$). If we take a look at the \outward binary, its initial eccentricity is $e \sim 0.14$, it increases to $\sim 0.2$ within a few kyr. This value remains nearly constant for a prolonged time and only starts increasing again at $t \sim 35$~kyr when another sink particle (S11) forms nearby. The eccentricity is affected by the external gravitational force until this new sink merges with one of the binary constituents at $t \sim 55$~kyr. Note that outward migration happens continuously over a long period of time (see Fig.~\ref{fig:torque_out}). This implies that the relationship between torques (except merger-induced torque) or the resulting outward migration and orbital (de-)excitation is insignificant. Interestingly, the periodic forcing from the accretion of gas of high-angular momentum at the apocentre (driving the outward migration) should circularise the orbit. However, we speculate that this does not happen because the periodic forcing happens only on the minidiscs while the forcing on the sinks by the gravitational torque is rather continuous over time (see Fig.~\ref{fig:disc_sink}). 

Finally, the inward-migrating binaries (middle panels) show a behaviour more in line with what is observed in normal protostellar discs and/or protoplanetary discs: the orbits that are initially eccentric circularise while the stars migrate inward, due to the loss of angular momentum.

\section{Cases without outward migration}
\label{sec:result2}

\begin{figure}
    \centering
	\includegraphics[width=0.48\textwidth]{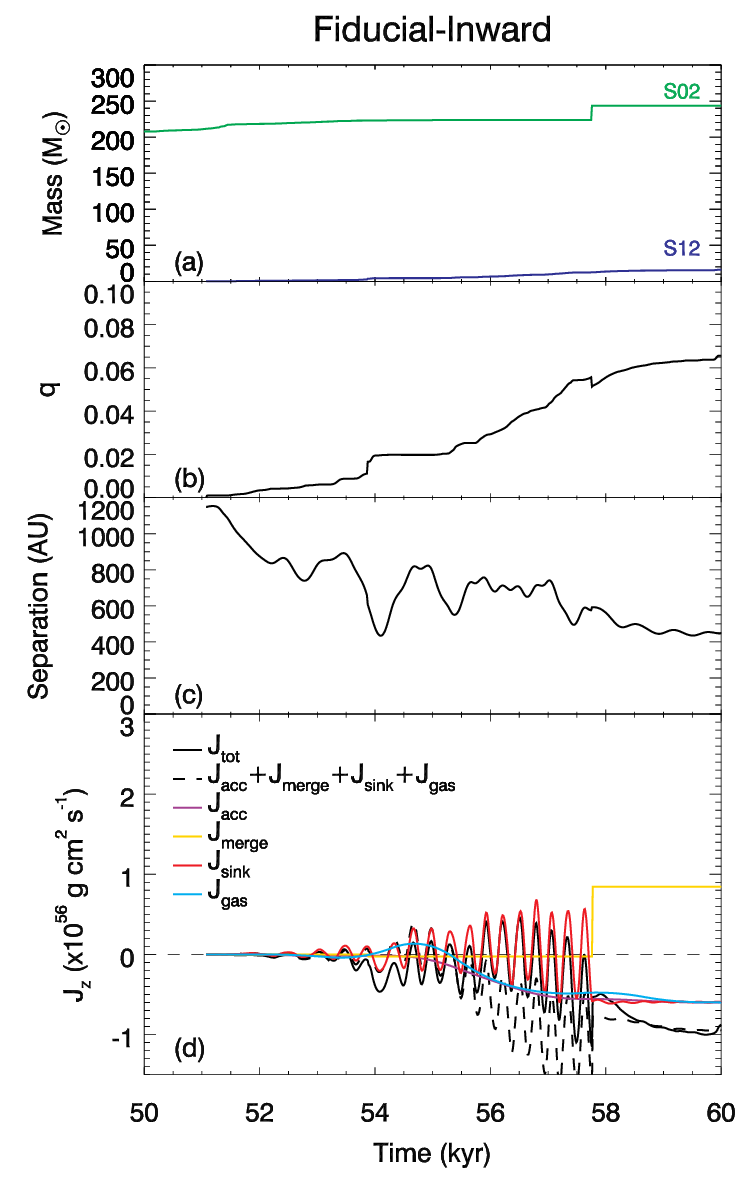}
    \caption{Same as Fig.~\ref{fig:torque_out} but for the \inward binary. The stars migrate inward (Panel c). Notable differences with respect to the \outward binary case are the small mass ratio (Panel a and b) and the negative gravitational torque from the gas disc (light blue, Panel d).}
    \label{fig:torque_in}
\end{figure}

\begin{figure}
    \centering
	\includegraphics[width=0.48\textwidth]{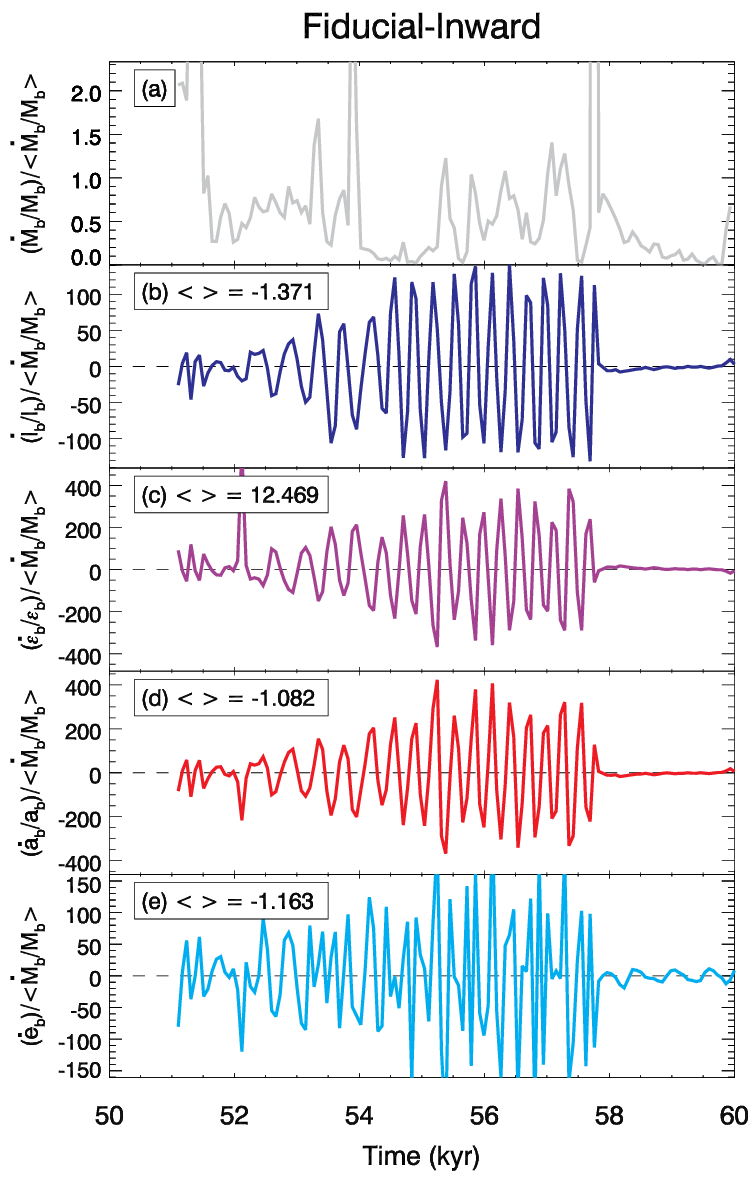}
    \caption{The time evolution of $\dot{M_{b}}/M_{b}$, $\dot{l_{b}}/l_{b}$, $\dot{\mathcal{E}_{b}}/\mathcal{E}_{b}$, $\dot{a_{b}}/a_{b}$, and $\dot{e_{b}}$ for the \inward system. We show only the first $\sim 10$~kyr during which the binary migrates inward. The format is the same as in Fig.~\ref{fig:energy_out}. The rates are normalised by $1/t_{acc} \approx 43$~kyr. The data is smoothed over 5 points ($\approx 0.07$~kyr). Contrary to the outward case, the angular momentum and semimajor axis (Panels b and d) decrease meaning the orbit shrinks.}
    \label{fig:energy_in}
\end{figure}

\begin{figure}
    \centering
	\includegraphics[width=0.48\textwidth]{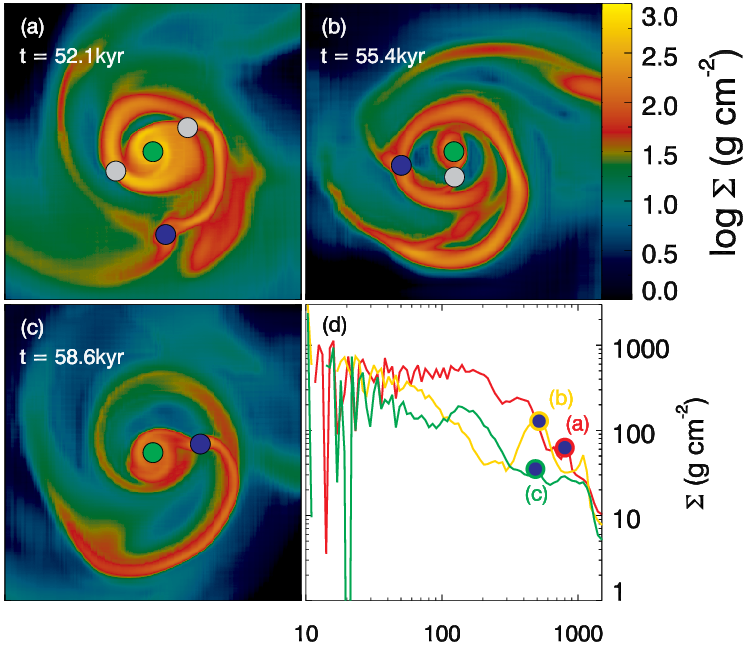}
    \caption{Panels (a) to (c): Snapshots of the gas surface density of the Fiducial-Inward binary system at different times. Each image is centered on the primary star (S02, green circle). The secondary S12 is shown as a blue filled circle and the other short-lived sinks are shown as grey circles. S12 forms in a spiral arm of the minidisc (a). As S12 migrates inward, it grows in mass and opens a gap (b). Once the inward migration stalls, the spiral structures exerting a negative torque disappear (c). Panel (d): Gas surface density profile in the three panels are shown with lines of different colors. The secondary star opens a gap at $t \sim 55$~kyr (yellow line). }
    \label{fig:snap_inward}
\end{figure}

Most binaries migrate outward in our simulations but there are a few exceptions (see runs labeled as "neutral" or "inward" in Table~\ref{tab:sim}). To better understand the conditions necessary for outward migration, we select some cases in which it does not happen and we make a comparative analysis. Fig.~\ref{fig:torque_in} shows the time evolution of the mass, mass ratio, separation, and angular momentum and torques for the case \inward, and can be compared directly, having the same format, to Fig.~\ref{fig:torque_out} for our \outward case. The simulation ends at $t \sim 92$~kyr but we show only the first $10$~kyr to focus on the initial inward migration. In this binary, similarly to the \outward case, the minidisc of the primary (S02) fragments to form the secondary star (S12). However, for this binary, the initial separation between the stars is initially about 2-3 times than for the \outward case ($1200$~AU) but decreases down to $400$~AU during the first $\sim 10$~kyr as S12 migrates inward (third panel). This inward migration is caused by the accretion (purple), gravitational torque from the gas disc (light blue line), and other stars in the system (red line, Panel~d). Similarly to the \outward case, the accretion torque is negative but, unlike in the outward case, the gravitational torque is also negative. The angular momentum evolution due to the gravitational force from sink particles (red line) oscillates because other sink particles formed together in the minidisc of S02. Before they merge ($t \lesssim 58$~kyr), however, the average value is much smaller than the other contributions to the angular momentum evolution, therefore forcing from other sink particles does not contribute to the inward migration substantially for the first several kyrs. At $t \gtrsim 58$~kyr, this term remains negative because sinks disappear through mergers and this angular momentum offset is compensated by the angular momentum term due to the mergers that becomes positive (yellow line). In Fig.~\ref{fig:energy_in} we show $\dot{M}_b$, $\dot{l}_b$, $\dot{\mathcal{E}}_b$, $\dot{a}_b$, and $\dot{e}_b$ as a function of time for the \inward binary. The figure confirms that, contrary to the \outward case, external forces extract orbital angular momentum and energy (Panel~b and c) causing the inward migration (Panel~d). During the inward migration, we notice that a prominent spiral structure develops and it disappears nearly when the migration stops. We speculate, therefore, that this spiral structure drives the inward migration as found in CH19. We present gas density maps for the Fiducial-Inward binary in Fig.~\ref{fig:snap_inward}. Non-axisymmetric spiral structures developed within the orbit of the secondary star (S12, blue circle) in the early stage (Panel~a) but this feature becomes less clear at a later stage (Panel~c) when the secondary opens a gap. In Section~\ref{sec:disc_out}, on the other hand, we argued that the fundamental reason for the outward migration is the accretion of high angular momentum gas by the circumstellar discs that migrate outward and drag the stars with them by gravitational torques. This implies that this latter process is weakened or suppressed in this binary. A possible explanation for this difference is the small mass ratio of the binary stars (Fig.~\ref{fig:torque_in}, Panel b). The mass ratio for this binary remains small ($q \lesssim 0.01$) for the first $\sim 10$~kyr when the inward migration takes place, also implying a lower gas accretion rate with respect to the \outward case. On the other hand, in the outward migration case in Fig.~\ref{fig:torque_out}, the secondary star grows quickly and retains mass comparable to the primary. This difference in mass ratio and accretion rate is critical to the direction of migration for the following reasons. First, as discussed in Section~\ref{sec:eff} and Fig.~\ref{fig:J}, the mass ratio of the stars determines the efficiency of the accretion of high angular momentum gas for a fixed mass accretion rate. With a small mass ratio, the secondary, which accretes a gas at the outer orbit, has nearly Keplerian angular momentum. In this case, the change in angular momentum through accretion is reduced and the binary cannot overcome the torque from the spiral structure. Second, if the mass ratio is large enough, the binary opens a gap more easily then the migration can turn into an outward one by accreting gas (CH19). 
The dependence of the gap opening on the binary mass ratio has been reported also in earlier studies \citep{escala2005,crida2006}. On the other hand, if the secondary star fails to gain a large mass and the mass ratio remains small, it cannot open a gap to halt the inward migration. In conclusion, the mass ratio of binary plays a crucial role in the direction of migration because it affects the efficiency of angular momentum accretion and gap opening. Note that the binaries with inward migration have large initial separations and small mass ratios (Table~\ref{tab:sim}). The secondary star cannot open a gap and does not migrate outward and grow in mass. If a binary, on the other hand, forms at the centre of a barred spiral structure, stars have a small initial separation and typically mass ratios closer to unity, thereby they can more easily start the outward migration that is self-sustained by continuously accreting higher angular momentum gas from the circumbinary disc. Our interpretation of mass ratio as the reason for inward migration is consistent with previous works. The binary shrinks for the first few kyrs when the mass ratio remains small and stops shrinking when $q \sim 0.05$. The orbit slowly expands later on as in CH19 while the binary keeps the mass ratio greater than this. This critical value of the mass ratio $q$ is consistent with the one suggested in the literature. For instance, \citet{munoz2020} and \citet{duffell2020} predicted that inward migration happens when $q \lesssim 0.2$ and $0.05$, respectively. In \citet{dempsey2021}, on the other hand, binaries migrate inward when the value is even lower ($q \lesssim 0.01$).

\begin{figure}
    \centering
	\includegraphics[width=0.48\textwidth]{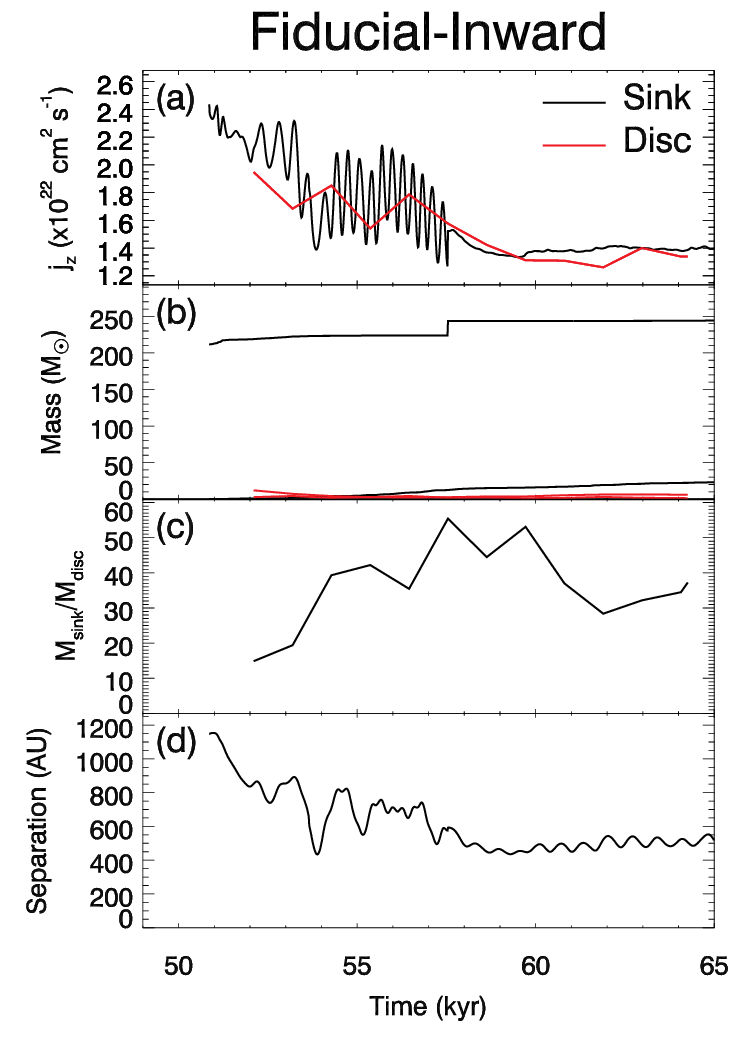}
    \caption{Same as Fig.~\ref{fig:disc_sink} but for the \inward binary case. The orbit initially shrinks and later the separation remains nearly constant. Note the disc mass is always negligible in this case and the secondary star forms at a wide separation from the primary.}
    \label{fig:disc_sink_inward}
\end{figure}

\begin{figure}
    \centering
	\includegraphics[width=0.48\textwidth]{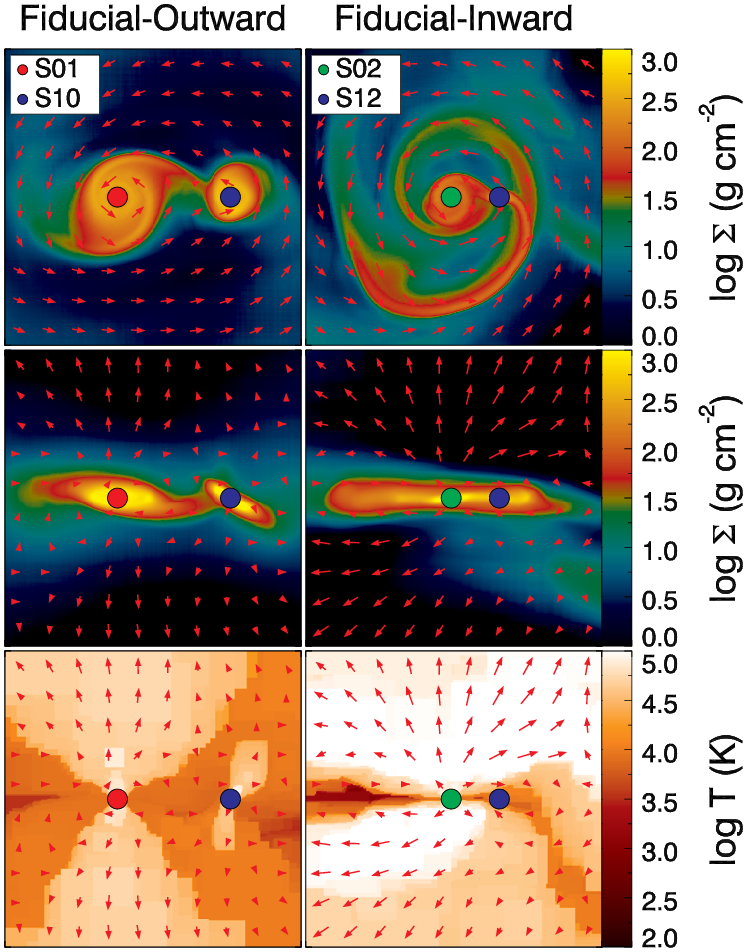}
    \caption{Snapshots of outward (left) and inward(right) migration binaries. The top two panels show the face-on and edge-on gas densities. The bottom panels show the gas temperature (edge-on). Sink particles are shown as circles and red arrows represent the velocity fields. Compared to \outward, \inward binary lacks a circumstellar disc of the secondary star (blue circle) due to the strong radiative feedback (bottom panels).}
    \label{fig:snapshot}
\end{figure}

Finally, we discuss another consequence of the mass ratio that is likely relevant: radiative feedback from the central massive stars. To motivate this effect, we briefly introduce one of the main results of Paper~II, which did not include radiative feedback from the protostars. According to simulation results in that paper, there is a correlation between the accretion rate, that is proportional to the gas temperature of the disc and regulated by the external irradiation from X-rays, and the maximum distance of the stars from the centre. This result suggests the high accretion rate causes the outward migration of the stars. In Paper~III, however, this trend was less clear. For instance, some binary stars in a strong X-ray background were found to expand with time to large distances ($d_{max} \sim $~8,000~AU, Run~E in Table~\ref{tab:sim}) while binaries in a moderate X-ray background (e.g., Run~B and Run~D) have smaller separations ($d \sim $~2,000~AU), even though the global accretion rate is lower in the former. Our interpretation is that this discrepancy between the results in Paper~II and Paper~III occurs because the radiative feedback from massive stars weakens the correlation between the accretion rate and outward migration found in Paper~II. In the feedback model used in Paper~III \citep{hosokawa2009,hosokawa2010}, the luminosity of a star is not a linear function of the mass and accretion rate. For this reason, the local accretion rate of the stars deviates from the global accretion trend regulated by an X-ray background. In addition, under strong protostellar feedback, the gas disc is evaporated and remains small compared to the stars (compare Panel c of Fig.~\ref{fig:disc_sink} and  Fig.~\ref{fig:disc_sink_inward}). The result is that the secondary star lacks a significant circumstellar minidisc which accretes high angular momentum gas and drags the star in the direction of the motion. With the lack of a minidisc, stars may not overcome the angular momentum loss by the inner spiral arms (CH19). The process is also visualised in Fig.~\ref{fig:snapshot}. In \inward case, the primary star (green circle) creates an intense UV radiation field (bottom panel) and the secondary (blue circle) lacks a prominent circumstellar minidisc. 

As seen in Table~\ref{tab:sim} we have four binaries without outward migration (inward and non-migrating). In these binaries, the primary stars are either born massive (\non) or the secondary star forms so late that the primary has enough time to grow in mass by accretion (\inward). Then the massive stars evaporate the gas discs and suppress gas accretion from the envelope. In this condition, it becomes harder for the small secondary to halt the inward migration and initiate the outward migration by accreting high angular momentum gas. A caveat of this work is that a small minidisc may be present even with strong radiative feedback but it may not be resolved in our simulations due to the insufficient sink resolution. Since the minidisc is poorly resolved or unresolved in cases with small mass ratios we cannot rule out that the inward migration is due to resolution effects and outward migration is possible in better-resolved simulations. Although the resolution effect cannot be completely ruled out, we speculate the mechanism for inward migration is still valid because the efficiency of angular momentum accretion (Section~\ref{sec:eff}) is not sensitive to the resolution.

\section{Discussion}
\label{sec:discussion}

We find that most massive Pop~III binary stars form at relatively close separations ($\sim 500$~AU) but migrate outward to form wide Pop~III binaries (5,000$-$10,000~AU). If these binaries survive with such orbits for the next $\sim 2$~Myr, when the stars explode, they become wide black hole binaries. Thanks to their wide separation, they have greater opportunities of being perturbed by field stars than close binaries. When perturbed, the orbits of the wide binaries are dynamically hardened \citep{liu2020}. Furthermore, if the orbits are eccentric (e.g., \non), the orbits may get excited and become more eccentric reducing their pericentre distance. If the distance at pericentre becomes sufficiently small, the emission of gravitational waves becomes important in removing angular momentum eventually leading to BHB merger and the emission of a strong GW signal within a Hubble time \citep{michaely2019,michaely2020}. 

The role of radiative feedback in migration has been discussed by many authors. This has been considered insignificant when stars migrate inward because the migration time scale is shorter than the Kelvin-Helmholtz time scale after which stars become luminous in the UV \citep{stacy2010,inayoshi2014,latif2015,hirano2017}. In addition, by modifying the gas profile around sink particles, \citet{li2022} reported heating of the minidiscs may change the direction of the migration suggesting radiative feedback may play a dominant role. \citet{del_valle2018}, on the other hand, implemented AGN wind and mechanical feedback to idealised CBD simulations and found that AGN feedback is important in the fast migration case. In this work, we did not perform a systematic numerical study focusing on the responses of the minidiscs to thermal/mechanical feedback and their impacts on migration as in previous works \citep{del_valle2018,li2022}. We found, however, that radiative feedback may contribute to migration in particular situations as discussed in Section~\ref{sec:result2}. The main difference between this work and others is the fact that binaries can form at late times. If the UV photons from the primary star are blocked in the disc plane, the minidisc of the primary is not fully evaporated and the secondary can form. At the same time, the radiative feedback from the primary is strong enough to prevent the secondary from growing. If the secondary is kept small in size, the gap opening is delayed and the star is found in a smaller orbit. Furthermore, the secondary competing with the massive companion cannot accrete high angular momentum gas and thus fails to migrate outward. We note, however, that the result is sensitive to the feedback prescription \citep{jaura2022}. We artificially reduce the amount of radiation in the disc plane (See Section~2 of Paper~III). If the feedback is strong enough to evaporate the disc quickly ($\sim 10$~kyr), outward migration would occur less frequently due to the lack of high angular momentum gas. We leave numerical experiments on the effects of changing the radiative feedback recipes on the binary migration in an idealised setup as future work.

In this paper, we do not analyse in depth the role of alignment/misalignment between the minidiscs and the CBD, but this may be potentially important for migration as pointed out by other authors before \citep{miranda2015,moody2019,tiede2024}. As seen in the middle left panel of Fig.~\ref{fig:snapshot}, the minidiscs of \outward are tilted with respect to the orbital plane (as well as the CBD). The fact that this binary with the misaligned discs migrates outward is consistent with the results in \citet{moody2019} but it is opposite to the main result in \citet{miranda2015}. We found that minidiscs are slightly tilted ($\lesssim 20^\circ$) in binaries migrating outward. However, studying the relation between tilt and migration is beyond the scope of this paper, and this topic should be revisited with a more systematic approach in future work.

An interesting discussion is a trend in migration across different metalicities. \citet{he2023} reported that metal-rich stars preferentially migrate inward. The difference from our work is that these systems are dominated by the central stars while the other stars have relatively low masses (that is, low mass ratio $q$). In this case, as mentioned in Section~\ref{sec:eff}, the angular momentum difference between the secondary star and the gas is reduced. Since the star accretes less angular momentum, the outward migration becomes less efficient. In our simulations, on the other hand, stars survive if they have masses comparable to the primary and migrate outward efficiently. 

Differences in gas metallicities may result in different behaviour of migration in the two populations of stars. A primordial gas cloud has a high temperature due to inefficient \hm\ cooling. Since the accretion rate is proportional to $c_{s}^3$, Pop~III star formation happens in a gas cloud with a high accretion rate. As seen in Paper~II, however, a disc in this environment is more Toomre-unstable and thus fragmentation is more active. Stars forming in this disc are likely to have similar masses and migrate outward in the end. On the contrary, the opposite happens in a metal-rich gas cloud. The accretion onto each prestellar core remains low and thus the disc is Toomre-stable and smooth as seen in \citet{he2023}. Once stars form, however, the evolution of the system is dominated by the massive central star and smaller secondary easily migrate inward as discussed in Section~\ref{sec:result2}. An interesting question is whether there is a critical metallicity above which stellar migration transitions from outward to inward migration. Since it is the accretion rate determining this transition, and the accretion rate is also responsible for the typical high-masses of Pop~III stars and their top-heavy IMF, it is likely that the critical metallicity is the same determining the transition from top-heavy to Salpeter IMF \citep[e.g.,][]{chon2021}.


\section{Summary}
\label{sec:summary}
We exploit radiative hydrodynamics simulations of star formation at the centre of metal-free minihaloes to explore the mechanisms leading to the outward migration of Pop~III protostar binaries. The initial conditions are created by extracting the central regions of two minihaloes irradiated by LW/X-ray backgrounds of various intensities as described in Paper~III. In each simulation, the typical outcome is the formation of hierarchical binaries or more generally multiple-star systems with a top-heavy IMF. The orbits of the binaries are typically elliptical and have a wide separation that increases with time, i.e., outward migration is a common occurrence. We investigate the time variation of the orbital angular momentum of the stars and the different torques and/or accretion of angular momentum leading to its increase. Below we summarise the key findings:
\begin{enumerate}
    \item Multiple stars form out of disc fragmentation. Some of them migrate inward and merge with the primary star on relatively short timescales, often producing an initial ellipticity significantly larger than zero of the stellar orbits of surviving stars. Of the stars that survive without merging for a longer time, however, most of them form binaries that migrate outward (for 10 out of 14 binary systems) producing wide binaries with separations up to 5,000$-$10,000~AU and elliptical orbits. The circumstellar minidiscs around each of the stars in this wide binary often fragment forming a quadruple hierarchical system in which even the binary stars in the minidiscs migrate outward.  We conclude that outward migration and the formation of wide Pop~III binaries with elliptical orbits is the most common outcome for Pop~III stars.

    \item Pop~III protostars obtain orbital angular momentum from the gravitational torque of their circumstellar minidiscs. These minidiscs, lose angular momentum to the stars but gain angular momentum by accreting high angular momentum gas from the circumbinary envelope. The accretion of angular momentum per unit mass is most effective in an equal-mass binary because their orbital velocity is significantly lower than that of the Keplerian disc at the same radial distance from the CoM. Then the angular momentum in the minidisc is transferred to the stars by gravitational torque and therefore the binary expands with time. On the other hand, inward migration happens in a binary that forms with a wider separation and with a small mass ratio ($q=m_2/m_1 \lesssim 0.1$), because in this case, the angular momentum accretion is less effective.

    \item Outward migration happens with the following periodic cycle. At the apocentre, the minidiscs grow in mass and migrate outward by accreting high angular momentum gas from the circumbinary disc. At pericentre, the accreted gas stored in the minidisc is accreted onto the stars. This cycle repeats until the density of the circumstellar minidiscs are low and the accretion rate of high angular momentum gas becomes negligible (at separations of 5,000$-$10,000~AU).
    
    \item Unlike previous studies, we find radiative feedback from protostars may affect the migration of stars in certain conditions. When a binary forms via late-time fragmentation, the massive primary star blows out the gas in the disc, suppressing the accretion rate on the secondary star and preventing it from gaining angular momentum by gas accretion. In these cases, the gravitational torque from spiral structures produces a net negative torque and therefore the binary star spirals inward.

    \item The lack of efficient cooling in a gas of primordial composition is the ultimate reason for the top-heavy IMF of Pop~III stars and their outward migration, which is not observed as frequently in Pop~II star formation \citet{he2023}. As the accretion rate is proportional to $c_{s}^3$, metal-free discs are relatively massive and Toomre-unstable, fragmenting rapidly. In these discs nearly equal-mass Pop~III binaries easily form in barred spiral structures at the centre of the disc, producing a configuration where disc gas with higher angular momentum than the stars is accreted most effectively. Pop~II star formation, on the other hand, is dominated by the central massive star and a disc with significantly less mass and Toomre-stable. According to \citet{he2023} the fragments in the disc are pre-existing and simply accreted into the disc from outside. In this case, the angular momentum accretion is less effective and therefore Pop~II stars tend to migrate inward and often get ejected by 3-body interactions.

    \item Binaries are typically born eccentric through 3-body encounters/mergers or non-axisymmetric potential. For outward migrating or non-migrating binaries the orbital eccentricity remains constant unless perturbed by mergers, while the orbits circularize for the fewer inward migrating cases. This implies that the periodic forcing at apocentre from torques producing outward migration do not excite or circularise the orbit. We speculate this is because the periodic force is applied onto the circumstellar discs while gravitational torque from the mini-disc to the star remains relatively constant over an orbital period.
\end{enumerate}

\section*{Acknowledgements}
We thank the anonymous referee for helping us improve our paper. The authors thank Dr. Takashi Hosokawa for kindly sharing the radiative feedback model of protostars. We make use of Deepthought2 and Zaratan operated by the University of Maryland (http://hpcc.umd.edu) to perform and analyse the simulations in this work.

\section*{Data Availability}

The data underlying this article will be shared on reasonable request to the corresponding author.



\bibliographystyle{mnras}
\bibliography{reference} 



\appendix

\section{Sink Particle Recipes}

The setup of our simulations differs from those of idealised circumbinary simulations widely used in the literature (\citealp{tang2017}; M19; \citealp{munoz2020,dempsey2020,dempsey2021,dittmann2021,dittmann2022}). To better understand the difference and estimate the potential impacts on our conclusion, we performed numerical tests and present the results in this appendix.

\subsection{Sink radius}
\label{app:rsink}

\begin{figure}
    \centering
	\includegraphics[width=0.48\textwidth]{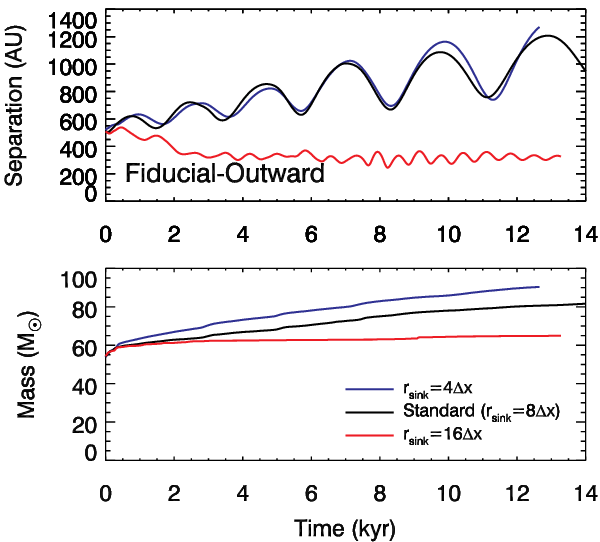}
    \caption{Separation (top) and total mass (bottom) of binaries with different sink radii. The fiducial value adopted in this work is $r_{sink}=8\Delta x$.}
    \label{fig:test_rsink}
\end{figure}

In previous works, the sink radii are kept smaller than $\lesssim 10$~\% of the semimajor axis (M19; \citealp{dittmann2021,dittmann2022,dittmann2023,sugimura2023}). The sink radius is an important parameter in a binary simulation as a too-large value might reduce the increase in the orbital angular momentum (M19). In this work, the choice of the sink radius is a more complicated matter because it determines the effect of radiative feedback. To better understand the effect of $r_{sink}$ on the orbital evolution, we present the test result with different sink radii in Fig.~\ref{fig:test_rsink}. When $r_{sink} = 4 \Delta x_{min}$ and $8 \Delta x_{min}$, the binary has similar orbits. With a larger sink radius ($r = 16 \Delta x_{min}$), however, the stars migrate inward. We speculate that feedback plays a significant role in this case. As can be seen in the mass plot (bottom panel), the growth of stars is suppressed significantly due to the effective radiative feedback. However, the possibility of the numerical effect (a large sink radius leads to an increase in angular momentum, M19) is not completely ruled out. Note that our sink resolution is sufficient for wide binaries ($a_{b} \gtrsim 2000$~AU) but is insufficient for close binaries. Therefore, the results of the close binaries must be interpreted with caution.

\subsection{Softening length}
\label{app:soft}

\begin{figure}
    \centering
	\includegraphics[width=0.48\textwidth]{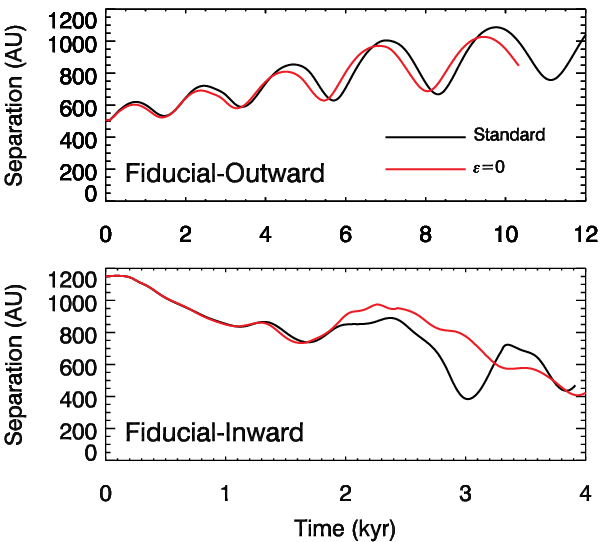}
    \caption{The time evolution of the separations of the \outward (top panel) and \inward binaries (bottom panel) with different assumptions for sink softening lengths. Each panel compares the binary separations for the fiducial simulation ($\varepsilon = 4\Delta x_{min}$, black) and the test run ($\varepsilon = 0$, red). Note that we kept the softening for gas and only changed the sink softening.}
    \label{fig:test_soft}
\end{figure}

In idealised CBD simulations, two sink particles are well-controlled, and therefore the codes typically do not require a softening term when calculating interactions between sink particles. The codes developed for cosmological simulations \citep[\ramses or \gadget,][]{teyssier2002,springel2005}, on the other hand, often handle N-body systems in which sink particles encounter other particles frequently and therefore these codes employ softening length. As we use \ramses in this study, we softened the gravitational force between two sink particles with the softening term $\varepsilon$ (equation~(\ref{eq:tau_star})). In this work, we choose $\varepsilon = 0.5 r_{sink} = 4 \Delta x_{min} = 50.4$~AU (Table~\ref{tab:sim}). In \inward, the minimum separation is $\sim 400$~AU and is only $\sim 8$~times larger than the sink radius. This causes $\sim 27$~\% difference in the strength of the gravitational force between sinks. In this work, any two sink particles closer than $202$~AU are not resolved and close binaries are beyond the scope of this paper. Rather, we focus on the formation of wide binaries and outward migration. In these binaries, the separations are even larger and the softening term becomes negligible compared to the separation. Therefore the effect of gravitational softening is insignificant in most cases. To confirm this, we present the result of the test run in Fig.~\ref{fig:test_soft}. We modified the code to keep the softening for gas accretion but erase the softening of the gravitational force due to other sink particles (i.e., $\varepsilon = 0$). In both \outward (top panel) and \inward (bottom panel), erasing the softening term (red lines) does not change the direction of the migration (outward to inward or vice versa). For \outward, the two orbits with different $\varepsilon$ do not match perfectly. Therefore, there is a possibility that other orbital parameters (such as semimajor axis or orbital period) may change over many periods ($\gtrsim 100$~orbits). However, gravitational softening does not change our conclusions on outward migration. We admit that inward migration is more sensitive to gravitational softening. As seen in the bottom panel, we argue that it does not affect the results of our two binaries with inward migration, but we need a follow-up study with high-resolution simulations.

\subsection{Accretion of angular momentum}
\label{app:accretion}

Recent numerical studies demonstrated the sink particle method with mass removal may introduce an artificial torque and thus the angular momentum of the accreted gas must be conserved to prevent this \citep{dempsey2020,dittmann2021}. We performed several test simulations and explored the impact of the accretion schemes on our conclusion. \ramses has a built-in option to turn on the angular momentum-conserving accretion prescription described above \citep[so-called `no-L' accretion,][]{bleuler2014}. This was devised to prevent the central sink from obtaining unphysically high angular momentum but this acts the same as in the works mentioned above.

\begin{figure}
    \centering
	\includegraphics[width=0.48\textwidth]{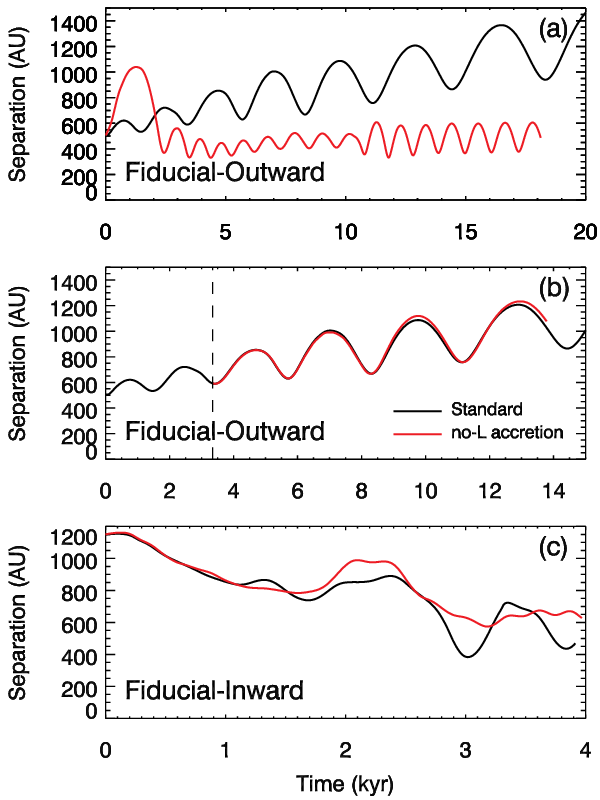}
    \caption{Numerical tests with different accretion schemes. In the standard method (black), the sink accretes the angular momentum of the gas in addition to its mass. In `no-L' accretion (red), on the other hand, the sink returns the tangential component of the velocity vector so that the angular momentum of the gas is conserved \citep{bleuler2014}. Panel a: Binary separation as a function of time for the Fiducial-Outward binary. The `no-L' option is turned on at when the binary forms ($t=0$~kyr). Panel b: Same as Panel a but the `no-L' accretion is turned on at $t\approx3.3$~kyr. Panel c: The binary separation for the Fiducial-Inward binary during the first $\sim 4$~kyr.}
    \label{fig:test_nol}
\end{figure}

In Fig.~\ref{fig:test_nol}, we present the results of numerical tests in three different situations. In each panel, we compare the fiducial simulation (black) and the one with `no-L' accretion (red). In the latter case, a sink particle returns the tangential component of the momentum to the accreted gas cells so that the angular momentum of the gas remains constant \citep[for details, see][]{bleuler2014}. In Panel~a, we compare the time evolution of the separation of the Fiducial-Outward binary using the two accretion methods aforementioned. With `no-L' accretion, the evolution in the first $2-3$~kyr differs significantly from the standard case. The secondary star has an eccentric orbit but is captured at the pericenter at $t \sim 2$~kyr and afterward remains in a close orbit ($\sim 400$~AU). Note that the prescription used in the simulation affects the initial gas properties substantially and this impact is highly unpredictable. Therefore, this test is inconclusive in determining whether the accretion scheme has a long-term effect on migration. To remove this initial chaotic effect due to complicated sink formation processes, we performed another test by turning on the `no-L' accretion at $t \sim 3$~kyr when the stars in the binary are well separated and after other sink particles disappear through mergers. The result is shown in Panel~b. Unlike in the previous case, the two orbits are similar. We speculate there are two main reasons why our result is different from what was found in previous works \citep{dempsey2020,dittmann2021}. First, the time scales of our simulations are much shorter than those in the aforementioned works. In our simulations, we followed the evolution of binaries for a few tens of orbits which correspond to the radiative feedback time scale \citep[$\sim 10 - 100$~kyr][]{hirano2014,hirano2015,sugimura2020}, while the works mentioned above focus on the long-term evolution of the binary. For instance, \citet{dittmann2021} evolved their binary for $\sim 2000$~orbits (see Fig.~2 of their work). If we run our simulations longer, hypothetically assuming that the gas supply is continuous, the change in orbital parameters may accumulate and become substantial. Second, the regulation of the density profile by the radiative feedback is likely dominant over effects related to assumptions on the angular momentum accretion scheme, unlike in the previous numerical studies that neglect radiative feedback effects \citep{dempsey2020,dittmann2021}. In Fiducial-Inward (Panel~c), the difference is insignificant and the binary shrinks irrespectively of the assumed accretion scheme. This is because the binary has a large initial separation ($\sim 1200$~AU) and thus is less sensitive to the accretion method.

Although different accretion schemes may cause numerical artifacts during the initial sink formation/disc fragmentation phase, our test confirms that the outward migration of Pop~III stars we found in this work is physical rather than numerical. This assertion is corroborated also by other evidence. First, outward migration is also found in other studies without mass removal and thus free from artificial torque (\citealp{chon2021}; Paper~I; Paper~II). In Paper~I and Paper~II we found that outward migration is common in Pop~III binaries. \citet{chon2021} focused on inward migration but found that Pop~III stars may migrate outward in certain situations. Secondly, \citet{he2023} used the sink particle method and found that metal-rich stars tend to migrate inward but outward migration is rare among them. We speculate this is due to the different gas properties mainly caused by the efficiency of cooling. If the artificial/numerical torque is the dominant process, they also should have found outward migration to be frequent. Finally, migration of Pop~III stars is found consistently in other studies with sink particle method \citep{sugimura2020,sugimura2023}. We do not know whether this `torque-free' accretion may play a role in orbital evolution for less massive systems, such as a planet orbiting a low-mass host star where radiative feedback is weaker. However, this effect is not dominant in massive Pop~III binaries because strong radiative feedback regulates the gas density profile and therefore does not change the direction of migration systematically.

\section{Non quasi-steady state accretion}
\label{app:aM}

\begin{figure}
    \centering
	\includegraphics[width=0.48\textwidth]{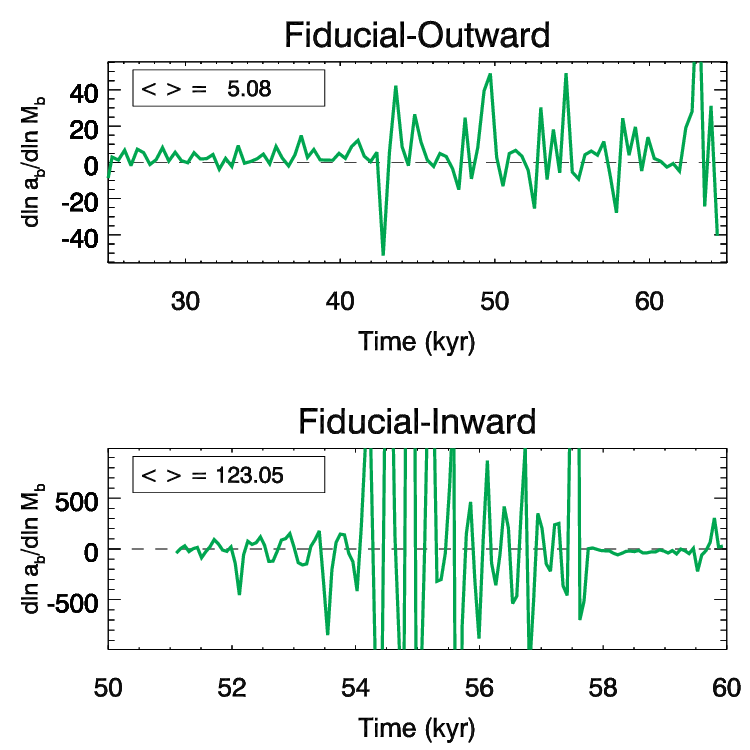}
    \caption{$\mathrm{d}\ln{a_{b}}/\mathrm{d}\ln{M_{b}}$ of \outward (top) and \inward (bottom). A large increase in the rate in the bottom panel is caused by a close encounter and merger between sink particles meaning that the disc is not quasi-steady state. }
    \label{fig:aM}
\end{figure}

In Fig.~\ref{fig:aM}, we plot $(\dot{a_{b}}/a_{b})/(\dot{M_{b}}/M_{b})$ which is also denoted by $(\mathrm{d}\ln{a_{b}})/(\mathrm{d}\ln{M_{b}})$ in the literature \citep{dittmann2021,dittmann2022}. In a quasi-steady state disc, this value is nearly constant \citep{lai2023}. In \outward (top panel), however, the rate is greater at later times. Before and after $t \sim 40$~kyr, the average rates are 2.27 and 4.21, respectively thereby suggesting the disc is not in a quasi-steady state like in typical CBD simulations. Note that the number of orbits covered in our simulations is smaller than the one required to reach the quasi-steady state ($\sim 100$~orbits, M19). In addition, the late-time evolution ($\sim 60$~kyr) is affected by the gravitational force of another sink particle (S11).

\section{Reference frame}
\label{app:frame}

Unlike in idealised simulations where a binary is fixed at the center of the box, binaries in our simulations drift and orbit the CoM of the entire system. The binaries are also accelerated by external forces but we did not consider this effect because the force on the CoM does not exert torque as seen below. Here, we will consider two reference frames: unprimed (unaccelerated) and primed (accelerated). The origins of both frames are at the CoM. In the unprimed frame, the force on the CoM is,
\begin{equation}
    M_{b}\boldsymbol{a} = m_{1}\boldsymbol{a}_{1} + m_{2}\boldsymbol{a}_{2} = \boldsymbol{F}_{1} + \boldsymbol{F}_{2} = \boldsymbol{F},
\end{equation}
where the subscripts mean individual sink particles, $\boldsymbol{a}$ is the acceleration, and $M_{b}=m_{1}+m_{2}$. The acceleration of Sink~$i$ in the primed frame is,
\begin{equation}
    \boldsymbol{a}'_{\subi} = \boldsymbol{a}_{\subi}-\boldsymbol{a}.
\end{equation}
Therefore, the force on the same sink is,
\begin{equation}
    \boldsymbol{F}'_{\subi} = m_{\subi}\boldsymbol{a}'_{\subi} = m_{\subi}\boldsymbol{a}_{\subi} - m_{\subi}\boldsymbol{a} = m_{\subi}\boldsymbol{a}_{\subi} - \frac{m_{\subi}}{M_{b}} \boldsymbol{F}.
\end{equation}
Finally, the torque on the CoM in the primed frame is,
\begin{equation}
    \begin{split}
        \boldsymbol{\tau}' &= \boldsymbol{r}_{1} \times \boldsymbol{F}'_{1} + \boldsymbol{r}_{2} \times \boldsymbol{F}'_{2} \\
        &= \boldsymbol{r}_{1} \times \left( \boldsymbol{F}_{1} - \frac{m_{1}}{M_{b}} \boldsymbol{F} \right) + \boldsymbol{r}_{2} \times \left( \boldsymbol{F}_{2} - \frac{m_{2}}{M_{b}} \boldsymbol{F} \right) \\
        &= \boldsymbol{r}_{1} \times \boldsymbol{F}_{1} + \boldsymbol{r}_{2} \times \boldsymbol{F}_{2} - \left( \frac{m_{1}}{M_{b}} \boldsymbol{r}_{1} \times \boldsymbol{F} + \frac{m_{2}}{M_{b}} \boldsymbol{r}_{2} \times \boldsymbol{F} \right) \\
        &= \boldsymbol{\tau} - \frac{1}{M_{b}} \left( m_{1} \boldsymbol{r}_{1} + m_{2} \boldsymbol{r}_{2} \right) \times \boldsymbol{F} \\
        &= \boldsymbol{\tau}.
    \end{split}
\end{equation}
Since the torques in the two frames are identical, we used the simple approach without subtracting force on the CoM.


\bsp	
\label{lastpage}
\end{document}